\documentclass{article}
\usepackage{psfig,amsmath,amssymb,a4}
\newcommand{\dt}{\partial_t}           
\newcommand{\dx}{\partial_x}           
\newcommand{\dz}{\partial_z}           
\newcommand{\grad}{\nabla}
\newcommand{\lap} {\nabla^2}
\newcommand{\ydot} {{\bf e_y} \cdot}
\newcommand{\geom} {q}
\newcommand{\expand}{\rho}
\newcommand{\coupa} {\beta}
\newcommand{\coupb} {\gamma}
\newcommand{\Coupa} {B}
\newcommand{\Coupb} {\Gamma}
\newcommand{\Tr}{\mbox{T}}
\newcommand{\Det}{\mbox{Det}}
\newcommand{\Disc}{\mbox{Disc}}
\newcommand{\sigmamid}{\sigma_{\rm mid}}
\newcommand{\Emid}{E_{\rm mid}}
\newcommand{\rt}{r_{\mbox{\tiny T}}}
\newcommand{\rc}{r_{\mbox{\tiny C}}}
\newcommand{\sigmat}{\sigma_{\mbox{\tiny T}}}
\newcommand{\sigmac}{\sigma_{\mbox{\tiny C}}}
\newcommand{\rmid}{r_{\rm mid}}
\newcommand{\rint}{r_{\rm int}}
\newcommand{\rpf}{r_{\mbox{\tiny PF}}}
\newcommand{\rh}{r_{\mbox{\tiny H}}}
\newcommand{\omegah}{\omega_{\mbox{\tiny H}}}
\newcommand{\rsn}{r_{\mbox{\tiny SN}}}
\newcommand{\Esn}{E_{\mbox{\tiny SN}}}
\newcommand{\rmidnl}{\tilde{r}_{\rm mid}}
\newcommand{\rintnl}{\tilde{r}_{\rm int}}
\newcommand{\Sgen}{\Delta}

\begin{document}
\numberwithin{equation}{section}
\title{Thermosolutal and binary fluid convection \\
as a $2 \times 2$ matrix problem}

\author{Laurette S. Tuckerman 
\vspace*{.2cm} \\
Laboratoire d'Informatique pour la M\'ecanique et les Sciences de
l'Ing\'enieur \\
BP 133, 91403 Orsay Cedex, France \\
email: laurette@limsi.fr}

\date{\today}
\maketitle
\vspace*{1cm}
\begin{abstract}

We describe an interpretation of convection in binary
fluid mixtures as a superposition of thermal and solutal
problems, with coupling due to advection and proportional
to the separation parameter $S$. Many of the properties of
binary fluid convection are then consequences of generic
properties of $2 \times 2$ matrices.
The eigenvalues of $2 \times 2$ matrices varying continuously
with a parameter $r$ undergo either {\it avoided crossing} or
{\it complex coalescence}, depending on the sign of the
coupling (product of off-diagonal terms).
We first consider the matrix governing the stability of 
the conductive state.
When the thermal and solutal gradients act in concert
($S>0$, avoided crossing), 
the growth rates of perturbations remain real and of 
either thermal or solutal type.
In contrast, when the thermal and solutal gradients
are of opposite signs ($S<0$, complex coalescence),
the growth rates become complex and are of mixed type.

Surprisingly, the kinetic energy of nonlinear steady
states is also governed by an eigenvalue problem
very similar to that governing the growth rates.
More precisely, there is a quantitative analogy between
the growth rates of the linear stability problem for 
infinite Prandtl number
and the amplitudes of steady states of the minimal
five-variable Veronis model for arbitrary Prandtl number.
For positive $S$, avoided crossing leads to a distinction
between low-amplitude solutal and high-amplitude thermal regimes.
For negative $S$, the transition between real and complex
eigenvalues leads to the creation of branches of finite
amplitude, i.e. to saddle-node bifurcations.
The codimension-two point at which the saddle-node bifurcations 
disappear, leading to a transition from subcritical to supercritical
pitchfork bifurcations, is exactly analogous to the Bogdanov 
codimension-two point at which the Hopf bifurcations disappear 
in the linear problem.

\end{abstract}



\newpage
\section{Introduction}

Convection due to two competing or cooperating effects is
realized in a number of different physical systems:
In the thermosolutal or thermohaline problem, vertical thermal 
and concentration gradients are both externally imposed. 
In convection in binary fluids with Soret effect, 
only the temperature gradient is imposed, but cross diffusion 
induces a concentration gradient with similar properties.
An electrically conducting magnetic fluid may be subjected
to a vertical or a horizontal magnetic field, the
fluid layer may be rotated, or two solutes may be introduced. 

In the 1960s and 1970s, Veronis \cite{Veronis65,Veronis68} and other researchers 
\cite{Sani,Nield,BainesGill,Caldwell,HurleJakeman,SchechterVelardePlatten,PlattenChavepeyer75,PlattenChavepeyer76,HuppertMoore,Knobloch80} 
recognized the variety of behavior manifested by double-diffusive convection;
comprehensive texts and reviews \cite{Chandra,Gershuni,PlattenLegros} 
were written on the subject.
One of the reasons for studying these double-diffusive systems 
is that all display a common basic set of phenomena.
Both stationary and oscillatory instabilities occur, 
in other words pitchfork and Hopf bifurcations, the curves intersecting 
at parameter combinations which can be analytically calculated, 
at least approximately.
In the 1980s, double-diffusive convection became the 
paradigm example in a renaissance in the study of bifurcation theory 
and dynamical systems, as attention was focused on precisely such 
intersections, re-interpreted as codimension-two points 
by Knobloch et al. \cite{KnoblochProctor} and then by 
Brand et al. \cite{Brand}.

There followed a divergence of efforts, roughly speaking between,
on the one hand, detailed and rigorous mathematical analysis of the 
temporal complexity of thermosolutal convection and related problems
in a small container, e.g. 
\cite{DaCosta,GuckenheimerKnobloch,KnoblochMooreToomreWeiss,Rucklidge,KnoblochProctorWeiss,Bergeon}, and, on the other hand, 
realistic and physical exploration of the
spatial complexity of the Soret and other problems in a large container
for negative
\cite{BrethertonSpiegel,Knobloch84,Knobloch85,CoulletFauveTirapegui,Walden,Knobloch86,Ahlers,Deane1,Cross86,AhlersLucke,Moses87,LinzLucke87,Deane2,KnoblochMoore88,Bensimon,KnoblochMoore90b,KnoblochMoore90,Riecke,Schopf,Swinney,BartenBig,BartenBig2,Dominguez,HollingerBuchelLucke,HollingerLucke98,HollingerLuckeMuller,LuckeBig}
and positive
\cite{Legal,Moses86,Ahlers,Silber,MullerLucke88,Knobloch89,LhostPlatten,KnoblochMoore90b,Moses91,CluneKnobloch,BartenBig,Dominguez,Bergeon,Jung,Huke} 
values of the separation ratio.
The first line of research has led to an understanding of the mechanisms
producing global bifurcations, period-doubling, and chaos in 
two-dimensional convection in a confined geometry.
The second line of research has resulted in bifurcation diagrams 
detailing the transitions in an extended periodic geometry, primarily 
between patterns of different spatio-temporal symmetry.
For negative separation ratios, transitions occur between
steady convection, standing waves, and traveling waves.
For positive separation ratio, the transitions are between
patterns of rolls and squares, and also between 
weakly and strongly convective regimes.

\begin{figure}[h]
\centerline{
\psfig{file=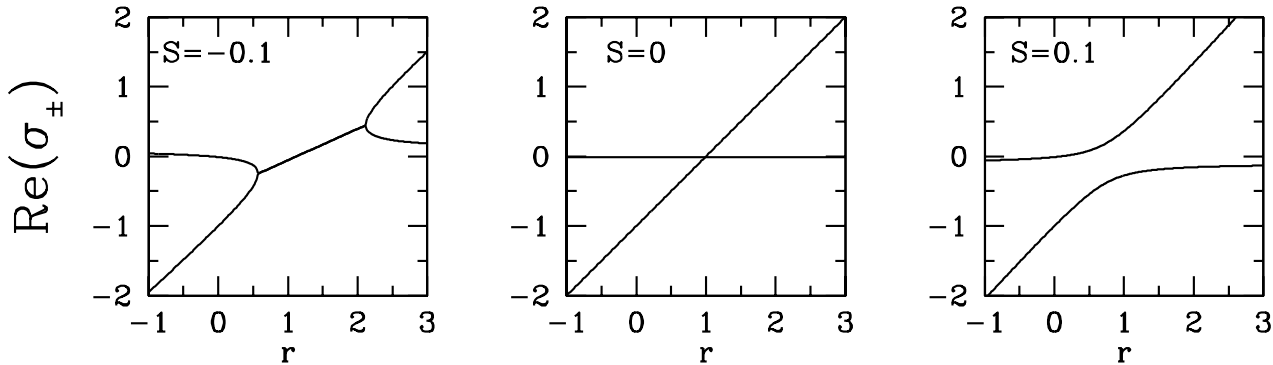}
}
\vspace*{-14cm}
\caption{{} 
Eigenvalues of thermosolutal convection 
as a function of scaled Rayleigh number $r$
for different signs of 
separation parameter $S$.
$S=-0.1$: formation of complex conjugate pair.
$S=0$: eigenvalues cross transversely. 
$S=0.1$: avoided crossing.}
\label{fig:eigenvalues}
\end{figure}

In the present article, we will use an analytically tractable
model of thermosolutal convection to examine the consequences of
a simple idea:
Consider a $2 \times 2$ real matrix
\begin{equation}
\left(\begin{array}{cc} \sigmat & \coupa \\ \coupb & \sigmac \end{array}\right)
\label{2x2}
\end{equation}
Its eigenvalues are
\begin{equation}
\sigma_\pm = \left({{\sigmat + \sigmac}\over 2}\right) \pm 
\sqrt{\left({\frac{\sigmat - \sigmac}{2}}\right)^2 + \coupa \coupb}
\end{equation}
Each of the elements $\sigmat$, $\sigmac$, $\coupa$, and $\coupb$
depend on parameters $r$ (scaled Rayleigh number) and 
$S$ (separation parameter). We assume that, for each $S$,
there is a value $\rint$ of $r$ at which $\sigmat$ and $\sigmac$ coincide.
We are interested in the behavior of the eigenvalues $\sigma_{\pm}$ in 
this vicinity.
There are three possibilities, depending on the sign of $\coupa\coupb$
at $\rint$.
If $\coupa\coupb < 0$, then $\sigma_+$ and $\sigma_-$ coalesce into a complex
conjugate pair in the vicinity of $\rint$.
If $\coupa\coupb = 0$, then $\sigma_+$ and $\sigma_-$ intersect transversely.
If $\coupa\coupb > 0$, then $\sigma_+$ and $\sigma_-$ are real and 
$\sigma_+ > \sigma_-$, with a change of slope at $\rint$.
This is the phenomenon known as avoided crossing.
These three possibilities, shown in figure \ref{fig:eigenvalues}, 
correspond to the three scenarios
observed when $S<0$, $S=0$, and $S>0$, respectively.
Moreover, we claim that this interpretation applies to the nonlinear
steady states as well as to the linear stability problem.
The remainder of the article is devoted to making these statements precise.

The paper is organized as follows.
Section 2 sets out the idealized free-slip thermosolutal problem that we will 
study. 
Although our calculations are limited to this analytically tractable case,
we will present evidence throughout the paper that our conclusions may
apply to double-diffusive convection problems in general.
Section 3 gives the classic linear stability analysis and the
bifurcations undergone by the system.
The behavior of the eigenvalues is interpreted
in the framework of avoided crossing vs. complex coalescence and 
also as primarily thermal vs. solutal.
Section 4 analyzes the minimal five-mode Veronis model.
We show that the steady states of the Veronis model obey a
two-variable eigenvalue problem entirely analogous to the
linear stability problem, with the energy playing the role of
the eigenvalue. We present the consequences of this analogy, 
particularly for the saddle-node bifurcations and 
codimension-two points that occur for negative $S$. 
For positive $S$, the transition between Soret and Rayleigh regimes 
is analyzed in detail as exemplifying the phenomenon of avoided crossing.
Section 5 sets out the simplest time-dependent system
reproducing the primary bifurcations and steady states
and, additionally, standing waves terminating in a heteroclinic orbit.
However, the traveling waves that have been the subject of so much
interest in double-diffusive convection are not accessible by 
our simplified approach; nor is the competition between patterns
of rolls and squares.
Section 6 is a brief summary and discussion.

\section{Thermosolutal problem}

We study the thermosolutal problem, in which thermal and concentration
gradients are imposed, instead of the more widely 
studied and experimentally accessible Soret-driven problem, in which
the concentration gradient results from the temperature gradient,
because its formulation is slightly simpler.
The basic phenomena which we will investigate occur in both 
problems, which have in fact been shown by Knobloch \cite{Knobloch80} 
to be formally equivalent for the idealized boundary conditions
we consider here: a two-dimensional geometry with free-slip upper 
and lower boundaries and imposed horizontal periodicity.
Although all our calculations concern the idealized
thermosolutal problem, we will also refer to results in
the literature derived using rigid boundary conditions or
concerning the Soret problem with various boundary conditions. 
We will not systematically refer to the related problems
of rotating or magnetoconvection (although these are also discussed in
\cite{Chandra,Gershuni,KnoblochProctor,GuckenheimerKnobloch,KnoblochProctorWeiss})
nor to the equally vast experimental literature
(for references, see, e.g. \cite{Schopf,BartenBig,BartenBig2,LuckeBig}).

In the thermosolutal problem, the flow is driven 
by vertical thermal and concentration differences $\Delta T$ and 
$\Delta C$ imposed across a layer of height $h$.
In the Boussinesq approximation,
the density is assumed constant except in the buoyancy term,
where it is taken to vary linearly with temperature and
concentration, with thermal and solutal expansion coefficients 
$\expand_T$ and $\expand_C$. 
There exists a steady motionless conductive solution,
with temperature and concentration profiles
depending linearly on the vertical coordinate $z$.

The stability of the conductive solution and the subsequent
evolution of the system
is determined by four nondimensional parameters.
The three diffusivities -- the thermal diffusivity $\kappa_T$,
the momentum diffusivity or kinematic viscosity $\nu$, 
and the solute diffusivity $\kappa_C$ -- are
described by two nondimensional ratios, chosen to be
the Prandtl number $P = \nu/\kappa_T$ 
and the Lewis number $L=\kappa_C/\kappa_T$.
For the linear problem, the Prandtl number is taken to be large;
$P = O(1)$ or $O(10)$ suffices.
The Lewis number is varied, but is considered to be small and,
for the figures, will be fixed at the frequently-studied value of $L=0.01$.
The thermal Rayleigh number 
$R \equiv g \expand_T \Delta T h^3/(\nu \kappa_T)$ measures
the imposed temperature gradient.
The concentration gradient can be specified by
an analogous solutal Rayleigh number, or by the separation 
parameter $S \equiv (\expand_C \Delta C)/(\expand_T \Delta T)$,
which may be either positive or negative.
Length, temperature, concentration, and time are 
nondimensionalized by $h$, $\Delta T$, $\Delta C$, and 
$h^2/\kappa_T$.

It will be convenient to write the equations governing the evolution 
of the solution in terms of both the vertical velocity $\hat{w}$ and
a velocity streamfunction $\hat{\phi}$, 
and temperature and concentration deviations $\hat{T}$ and $\hat{C}$ from 
their conductive profiles:
\begin{subequations}
\label{thermosolutal_full}
\begin{eqnarray}
\dt \hat{T} &=& \lap \hat{T} + \hat{w} 
+ \ydot (\grad \hat{\phi} \times \grad \hat{T})
\label{tsfT}\\
\dt \hat{C} &=& L \lap \hat{C} + \hat{w} 
+\ydot (\grad \hat{\phi} \times \grad \hat{C}) 
\label{tsfC}\\
\dt \lap \hat{w} &=& P\nabla^4 \hat{w} + PR\dx^2 (\hat{T} + S\hat{C}) 
+ \ydot \dx (\grad \hat{\phi} \times \grad \lap \hat{\phi})
\label{tsfphi} \\
\hat{w} &=& \dx\hat{\phi}
\label{tsw}
\end{eqnarray}
\end{subequations}
The boundary conditions are:
\begin{subequations}
\label{bcs}
\begin{eqnarray}
\hat{T} = \hat{C} = \hat{\phi} = 
\dz^2 \hat{\phi} = \hat{w} = \dz^2 \hat{w} &&\mbox{~~~~at~ $z=0,1$} \\
\hat{T}, \hat{C}, \hat{\phi}, \hat{w} 
&&\mbox{~~~~$2\pi/k$-periodic in $x$}
\end{eqnarray}
\end{subequations}
where $k$ may be fixed arbitrarily, or at the well-known value 
$k_{\rm crit} = \pi/\sqrt{2}$ which minimizes the convection threshold
in the case of free-slip boundaries.

Following, e.g. \cite{BainesGill,Knobloch80}, 
we additionally scale time and, consequently, velocity, by:
\begin{subequations}
\begin{equation}
\geom^2 \equiv k^2 + \pi^2 \label{geomdef}
\end{equation}
and introduce the geometrically scaled Rayleigh number:
\begin{equation}
r\equiv Rk^2/\geom^6 \label{rdef}
\end{equation}
\end{subequations}
Equations (\ref{thermosolutal_full}) become:
\begin{subequations}
\label{thermosolutal_scaled}
\begin{eqnarray}
\dt \hat{T} &=& \geom^{-2}\lap \hat{T} + \hat{w}
+ \ydot (\grad \hat{\phi} \times \grad \hat{T}) 
\label{tsfT_scaled}\\
\dt \hat{C} &=& L \geom^{-2}\lap \hat{C} + \hat{w}
+\ydot (\grad \hat{\phi} \times \grad \hat{C}) 
\label{tsfC_scaled}\\
\dt \geom^{-2}\lap \hat{w} &=& P\geom^{-4}\nabla^4 \hat{w} 
+ Prk^{-2}\dx^2 (\hat{T} + S\hat{C}) 
+ \geom^{-2}\ydot \dx(\grad \hat{\phi} \times \grad \lap \hat{\phi})
\label{tsfphi_scaled} \\
\hat{w} &=& \dx\hat{\phi}
\end{eqnarray}
\end{subequations}
\section{Linear analysis}\label{Linear analysis}
\subsection{Linear stability problem}\label{Linear_stability_problem}
We begin by discussing the linear stability of the conductive solution, governed by:
\begin{subequations}
\label{thermosolutal_lin}
\begin{eqnarray}
\dt \hat{T} &=& \geom^{-2}\lap \hat{T} + \hat{w} \label{tslT}\\
\dt \hat{C} &=& L \geom^{-2}\lap \hat{C} + \hat{w} \label{tslC}\\
\dt \geom^{-2}\lap \hat{w} &=& P\geom^{-4}\nabla^4 \hat{w} 
+ Prk^{-2}\dx^2 (\hat{T} + S\hat{C})
\label{tslphi}
\end{eqnarray}
\end{subequations}

Solutions to (\ref{thermosolutal_lin}) with boundary conditions (\ref{bcs}) 
are of the form:
\vspace*{-.5cm}
\begin{eqnarray}
\nonumber\label{solform}
\end{eqnarray}
\begin{displaymath}
\;\;\;\;\;\;\;\;\;\;\;\;\;\;\;\;\;\;\;\;\;\;\;\;\;\;\;\;\;\;
\;\;\;\;\;\;\;\;\;\;\;\;\;\;\;\;\;\;\;\;\;\;\;\;\;\;\;\;\;\;
\left(\begin{array}{c}\hat{T}(x,z,t)\\\hat{C}(x,z,t)\\\hat{w}(x,z,t)
\end{array}\right) =
\left(\begin{array}{c}T(t)\\C(t)\\w(t)\end{array}\right) 
\cos (kx) \sin(\pi z)
\;\;\;\;\;\;\;\;\;\;\;\;\;\;\;\;\;\;\;\;\;\;\;\;\;\;\;\;\;\;
\begin{array}{c}(\ref{solform}{\rm a})\\(\ref{solform}{\rm b})\\
(\ref{solform}{\rm c})\end{array}
\end{displaymath}
\addtocounter{equation}{1}
The time dependence of the linear system (\ref{thermosolutal_lin}) is:
\begin{equation}
\left( \begin{array}{c}T(t) \\ C(t) \\ w(t) \end{array} \right) 
=\exp(\lambda t)
\left( \begin{array}{c} T \\ C \\ w \end{array} \right)
\label{solformtime}\end{equation}
where $T, C, w$ are scalars denoting the amplitudes of the corresponding fields.
Substituting (\ref{solform})-(\ref{solformtime}) and (\ref{geomdef}) 
into (\ref{thermosolutal_lin}) yields the eigenvalue problem:
\begin{equation}
\lambda\left( \begin{array} {c} T \\  C \\ w \end{array} \right) 
=
\left(\begin{array}{ccc}
-1 & 0 & 1 \\
0 & -L & 1\\
Pr & PSr & -P\\
\end{array} \right)
\left(\begin{array}{c} T \\ C \\ w \end{array}\right) 
\label{thermosolutalmatrix}
\end{equation}

We simplify further by assuming that the Prandtl number $P$ is infinite.
Most quantitative results of interest to us depend only very weakly
on $P$ as long as $P \gtrsim 1$, e.g. \cite{Schopf}.
(Another approach to reducing the $3 \times 3$ problem to a $2 \times 2$
problem, which does not rely on $P$ large, is given in Appendix
\ref{Finite Prandtl number model}.)
The velocity amplitude $w$ is then related to $T$ and $C$ by the algebraic equation:
\begin{equation}
w =  r(T+SC)
\label{velslave}\end{equation}

After eliminating the velocity via (\ref{velslave}), the eigenvalue problem
(\ref{thermosolutalmatrix}) becomes:
\begin{equation}
\lambda\left( \begin{array} {c} T \\ C \\ \end{array} \right) 
=\left(\begin{array}{cc} r-1 & r S \\ r & r S - L\end{array}\right)
\left( \begin{array} {c} T \\ C \\ \end{array} \right) 
\label{lin2Dsys}\end{equation}
The matrix:
\begin{eqnarray}
\left(\begin{array}{cc} r-1 & r S \\ r & r S - L\end{array}\right) &=&
\left(\begin{array}{cc} -1 & 0 \\ 0 & -L  \end{array}\right) 
+r\left(\begin{array}{cc} 1 & S \\ 1 & S \end{array}\right) \label{linmatrix}\\
M &=& M_0 + r M_1 \nonumber
\end{eqnarray}
is of the form discussed in the introduction.
The matrix $M_0$ describes diffusion and the matrix $M_1$ describes advection
via (\ref{velslave}).
The diagonal elements of $M$:
\begin{subequations}\begin{eqnarray}
\sigmat & \equiv & r-1 \label{sigT}\\
\sigmac & \equiv & Sr-L \label{sigC}
\end{eqnarray}
can be viewed as the eigenvalues of a pure thermal and a pure solutal problem,
with a coupling of:
\begin{equation}
\coupa\coupb = Sr^2
\label{Sr2}\end{equation}\end{subequations}

The pure thermal convection problem for infinite Prandtl number
is obtained from (\ref{lin2Dsys}) by setting $S=0$.
Bifurcation to steady thermal convection occurs when $\sigmat = 0$, i.e. at:
\begin{subequations}\begin{equation}
r = \rt = 1 
\end{equation}

The pure solutal problem corresponds to convection driven exclusively
by concentration gradients, i.e. incomplete mixing of the
two species in the binary fluid. 
In the thermosolutal problem, the imposed concentration gradient
is set to the value $Sr$, while in the Soret problem, the
Soret coefficient $S$ is a property of the fluid.
Although we are studying the thermosolutal problem,
it will be convenient for us to consider $S$ as fixed and
to vary the single control parameter $r$ for all three problems:
thermal, solutal, and thermosolutal.
(In particular, all three problems have the same critical wavenumber
for the idealized boundary conditions.)
We therefore interpret (\ref{sigC}) as meaning that $S$ is fixed 
and the onset of convection occurs at:
\begin{equation}
r = \rc \equiv L/S
\end{equation}\end{subequations}
There are two cases, depending on the sign of $S$
($L$, the ratio of two diffusion coefficients, is always positive):
If $S > 0$, then $\sigmac(r)$ has positive slope and the conductive state 
is unstable for $r > L/S > 0$.
If $S < 0$, then $\sigmac(r)$ has negative slope and the conductive state 
is unstable for $r < L/S < 0$.

\subsection{Bifurcations}\label{Bifurcations_linear}

We now return to the coupled thermosolutal system (\ref{lin2Dsys}).
We begin by giving some exact results concerning the eigenvalues
and bifurcations of (\ref{lin2Dsys}).
Most of these results are well known, but we derive them here 
to illustrate our geometric interpretation and to prepare the
analogy with the nonlinear problem of section \ref{Nonlinear analysis}.
The results are summarized in the two large figures \ref{fig:linminus}
and \ref{fig:linplus}.

The uncoupled thermal and solutal eigenvalues (\ref{sigT}) and (\ref{sigC})
intersect at:
\begin{equation}
\rint=\frac{1-L}{1-S} 
\label{rint}\end{equation}
The behavior of the thermosolutal eigenvalues
of (\ref{lin2Dsys}) near $\rint$ is determined by the sign of 
the coupling strength (product of the off-diagonal terms) $Sr^2$,
which is in turn determined by the sign of $S$.
For $S$ positive, the eigenvalues remain real for all $r$, with 
avoided crossing near $\rint$. For $S$ negative, the eigenvalues
form a complex conjugate pair over an interval surrounding $r_{\rm int}$.

The eigenvalues of (\ref{lin2Dsys}) are:
\begin{subequations}
\begin{eqnarray}
\lambda_{\pm}
&=& {{\sigmat+\sigmac}\over 2} \pm 
\sqrt{\left({{\sigmat-\sigmac}\over 2}\right)^2 + Sr^2} \\
&=& {1\over 2} \left[(1+S)r -(1+L) \right] \nonumber\\
&&\;\;\;\;\ \pm {1\over 2}\sqrt{(1+S)^2 r^2 - 2(1-S)(1-L)r + (1-L)^2} \label{2Deigs2}\\
&\equiv& f(r) \pm \sqrt{g(r)} \label{2Deigs3} 
\end{eqnarray}\label{2Deigs}
\end{subequations}
The linear function $f(r)$ and the quadratic function $g(r)$
are half the trace and a quarter of the discriminant, respectively, of the
matrix $M$ of (\ref{linmatrix}).
The qualitative behavior of (\ref{2Deigs}) depends on 
whether $g$ is positive for all $r$ ($S > 0$) or of both signs ($S<0$), 
and on whether $g$ is a quadratic function of $r$ ($S\neq -1$)
or linear ($S=-1$).
We write $\lambda_{\pm} = \sigma_{\pm} \pm i\omega$ to denote
the real and imaginary parts of $\lambda$, with
$\sigma_+=\sigma_-$ if $\omega\neq 0$.

Appendix \ref{Conic sections and eigenvalues}
derives properties of (\ref{2Deigs}) based on the matrices
$M_0$ and $M_1$ in (\ref{linmatrix}). 
Assuming that $S\neq -1$, then the discriminant $\Disc_1 = (1+S)^2$
of matrix $M_1$ is positive,
and so the real parts $\sigma_\pm(r)$ of (\ref{2Deigs}) describe a 
hyperbola given by:
\begin{eqnarray}
\left(\sigma + \frac{S+L}{1+S} - (1+S)
\left(r - \frac{(1-L)(1-S)}{(1+S)^2}\right)\right) 
\left(\sigma + \frac{S+L}{1+S}\right) &=& S\;\frac{(1-L)^2}{(1+S)^2}
\nonumber\\
\label{binhyp}\end{eqnarray}
and the imaginary parts $\pm \omega$ describe an ellipse given by:
\begin{equation}
\omega^2 + \frac{(1+S)^2}{4} \left(r - \frac{(1-L)(1-S)}{(1+S)^2} \right)^2
       = -S \frac{(1-L)^2}{(1+S)^2}
\label{binell}\end{equation}
(In the exceptional case $S=-1$, when $f(r)$ is constant and 
$g(r)$ is a linear function, then the curves $\sigma_\pm(r)$ and 
$\pm \omega(r)$ are parabolas; see figure \ref{fig:linminus}.)

The crucial quantity:
\begin{equation}
\Sgen \equiv S\frac{(1-L)^2}{(1+S)^2} 
\end{equation}
on the right-hand-side of both (\ref{binhyp}) and (\ref{binell}),
an invariant for second-degree equations 
(see Appendix \ref{Conic sections and eigenvalues} and \cite{Ayre}),
distinguishes between {\it avoided crossing} and
{\it complex coalescence}.
The sign of $\Sgen$ is determined by that of $S$.
If $S>0$, the eigenvalues remain real: the branches $\sigma_+$ 
and $\sigma_-$ remain distinct and continuous over all values of $r$.
If $S<0$, the branches $\sigma_{\pm}$ coalesce and there exists
an interval of $r$ over which the eigenvalues are complex.
$S$ also plays a role in determining parameters in 
(\ref{binhyp})-(\ref{binell}) other than $\Delta$, e.g.
the values ($\rmid$, $\sigmamid$) giving
the intersection point of the asymptotes of the
hyperbola or the center of the ellipse.
But it is the dependence of $\Delta$ on $S$ which
leads to the most striking results.

We now discuss each of (\ref{binhyp}) and (\ref{binell}) in turn.
Expression (\ref{binhyp}) can be inverted as follows:
\begin{equation}
r = {{\sigma_{\pm}^2
+\sigma_{\pm}(1+L) + L} \over
{\sigma_{\pm}(1+S) + (S+L)}} \label{rfcnl}
\end{equation}
(See e.g., \cite{Veronis65,Sani} for similar formulas.)
Equation (\ref{rfcnl}) gives $r$ as a single-valued function of $\sigma$,
despite the fact that (\ref{2Deigs}) would normally yield an equation
quadratic in both $\sigma$ and $r$.
Geometrically, this can be understood as follows 
(see Appendix \ref{Conic sections and eigenvalues}).
The roots of the two factors on the left hand side of (\ref{binhyp}) 
are the asymptotes of the hyperbola whose slopes in the $(r,\sigma)$ plane 
are the eigenvalues $\lambda_{1\pm}$ of the matrix $M_1$ in (\ref{linmatrix})
with determinant $\Det_1$.
Here, $\lambda_{1+}=1+S$, $\lambda_{1-}=0$, and $\Det_1 = 0$.
Physically, $\lambda_{1-}$ and $\Det_1$ vanish because the velocity $w$ 
advects both the conductive temperature and concentration profiles in 
the same way. Hence $\sigma = -(S+L)/(1+S)$ is a horizontal asymptote.
A line in the $(r,\sigma)$ plane which is parallel but not equal 
to an asymptote intersects the hyperbola in exactly one point 
(see Appendix \ref{Conic sections and eigenvalues}).
The consequence of this is that all real values
(except $-(S+L)/(1+S)$) of $\sigma$ are achieved exactly once.
(This distinctive feature of the eigenvalues arising from this problem
will also have repercussions on the nonlinear problem, discussed in
section \ref{Nonlinear analysis}.)
In particular (except when $S=-L$), $\sigma=0$ for a unique value of $r$,
\begin{equation}
\rpf = {L \over{S+L}} \label{rpf}
\end{equation}
which is the location of the unique steady bifurcation from the basic state,
well-known to be a pitchfork bifurcation.
For $S > (<) -L$, the steady bifurcation occurs at positive (negative) $r$,
with $\rpf \rightarrow +(-) \infty$ as $S \downarrow (\uparrow) -L$.

If $S>-L^2$ or if $S<-L$, then it is the upper branch $\sigma_+$ which crosses 
zero at the bifurcation, i.e. $\rpf$ satisfies $f+\sqrt{g}=0$ 
(see cases $S=-0.01, -0.1, -1$ of figure \ref{fig:linminus}).
If $S$ is in the range $-L < S < -L^2$, then $\sigma_-$ crosses zero,
i.e. $\rpf$ satisfies $f-\sqrt{g}=0$ (see case $S=-0.001$ of 
figure \ref{fig:linminus}).
At the endpoint $S=-L$ of this range, the steady bifurcation
goes to infinity, while at the other endpoint $-L^2$,
it coalesces with a Hopf bifurcation (see below).

We now consider the ellipse (\ref{binell}) describing $\omega(r)$.
If $S<0$, the eigenvalues are complex over the range:
\begin{equation}
\frac{1-L}{(1+\sqrt{-S})^2}
\equiv r_{-} < r < r_{+} \equiv
\frac{1-L}{(1-\sqrt{-S})^2}
\label{complex_range_lin}\end{equation}
The endpoints $r_{\pm}$ are solutions to $g=0$.
In this range, $\sigma_{\pm}$ is equal to half the trace
of the matrix $M$: 
\begin{equation}
\sigma_+ = \sigma_- = \frac{1}{2} \left[r-1 + S(r-L)\right]
\label{sigmalin}\end{equation}
A Hopf bifurcation occurs at $r=\rh$ if $\sigma_\pm = 0$ within the
range (\ref{complex_range_lin}) of complex eigenvalues.
Equations (\ref{sigmalin}) and (\ref{binell}) show that 
$\rh$ and $\omegah\equiv\omega(\rh)$ satisfy:
\begin{equation}
\rh = {{1+L}\over{1+S}} \label{rh} \;\;\;\;\;\;\;\;\;\;\;\;
\omegah^2 = -\frac{S+L^2}{1+S} 
\end{equation}
Thus, a Hopf bifurcation occurs if and only if $\rh$
falls within the range (\ref{complex_range_lin}), i.e.
\begin{equation}
-1 < S < -L^2
\label{hopfrange}\end{equation} 
If $S=-1$, then $\rh$, $r_+$, and $\omegah$ all become infinite.
This is the exceptional case in which $r(\sigma)$ forms a 
leftward-opening parabola and $r(\omega)$ a rightward-opening parabola
(see figure \ref{fig:linminus}). If: 
\begin{equation}
S=S_*=-L^2
\label{codim2lin}\end{equation}
\vspace*{-.2cm}
then:
\vspace*{-.2cm}
\begin{equation}
\rh = r_+ = \rpf = \frac{1}{1-L} \equiv r_*
\end{equation}
and $\omegah$ vanishes.
This is the well-known codimension-two point, e.g. 
\cite{Sani,Nield,BainesGill,HurleJakeman,SchechterVelardePlatten,KnoblochProctor,Brand,Knobloch86}; see figure \ref{fig:linminus}.

A detailed representation of the bifurcations undergone
by the thermosolutal system, made possible by logarithmic scaling,
is given in figures \ref{fig:linminus} (for negative $S$) 
and \ref{fig:linplus} (for positive $S$).

\subsection{Linear thermal and solutal regimes}
\label{Linear thermal and solutal regimes}
\begin{figure}
\vspace*{-4cm}
\centerline{
\psfig{file=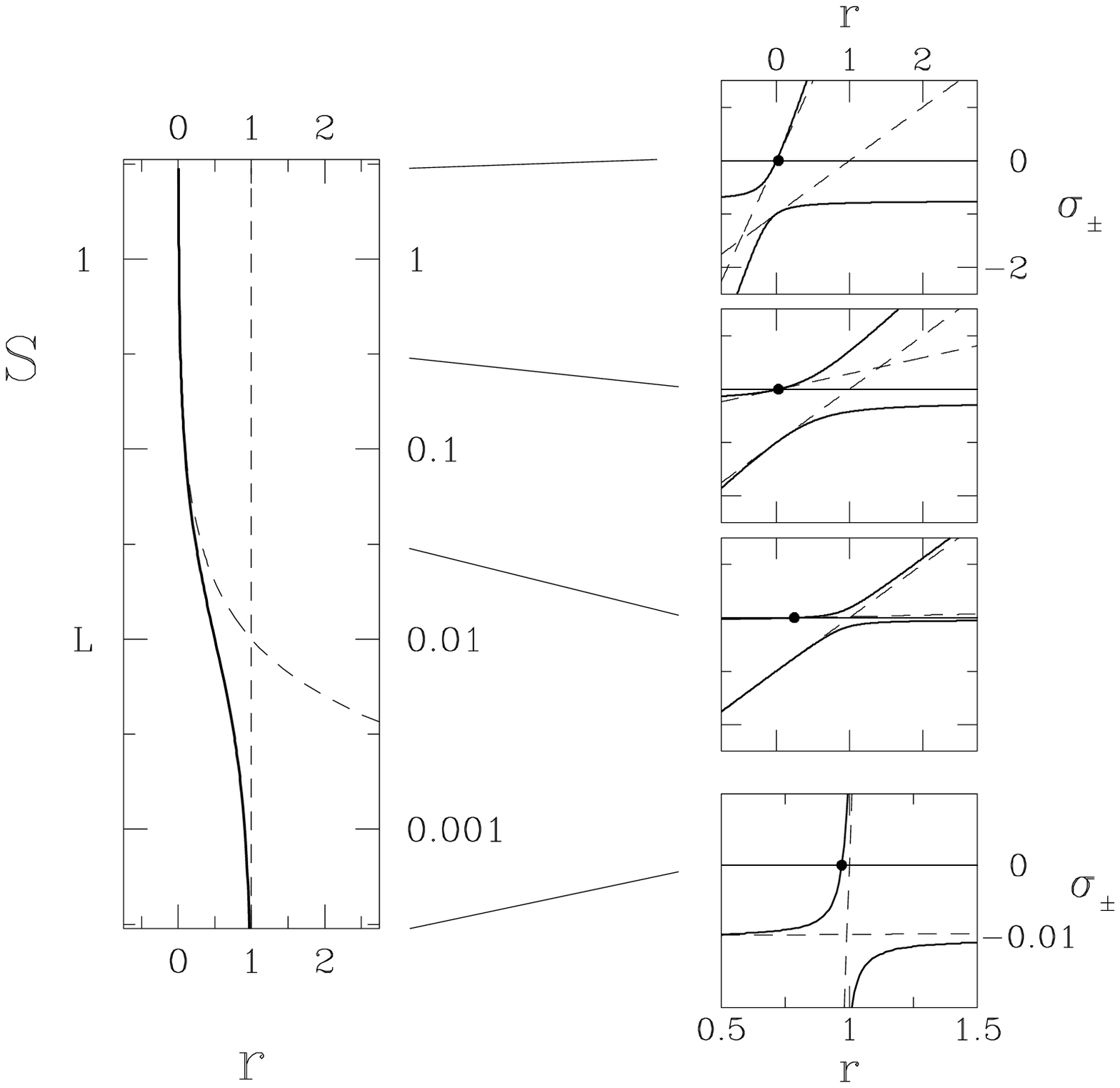,width=20cm}}
\caption{{}
Behavior of thermosolutal eigenvalues for positive $S$.\newline
Left: thresholds for positive values of $S$ plotted on a logarithmic scale.
Solid curve shows the threshold $\rpf=L/(L+S)$ of steady bifurcations. 
Dashed curves show the thresholds $\rt=1$ and $\rc=L/S$ of the pure 
thermal and solutal problems, which are within 0.1 of $\rpf$ for
$S\lesssim 0.1 L = 0.001$ and $S\gtrsim L/\sqrt{0.1} = 0.03$,
respectively. \newline
Right: Real part $\sigma_\pm$ of the eigenvalues of the thermosolutal
problem as a function of $r$ for representative positive values of $S$.
Dashed lines are the eigenvalues $\sigmat$ and $\sigmac$ of the pure 
thermal and solutal problems, with slopes 1 and $S$, respectively
and intersection point $\rint$. Dots represent bifurcations.
$S=0.0003$ illustrates an extreme case of avoided crossing:
$\sigma_+$ is very close to $\sigmac$ for $r < \rint$ 
and to $\sigmat$ for $r > \rint$.
This is in the thermal-dominated regime:
$\rpf=0.97 \approx 1 = \rt$.
As $S$ increases, $\sigma_\pm$ deviate more from the pure eigenvalues.
For $S=0.03$, $\rpf=0.25$, as compared to $\rt=1$ and $\rc=0.33$.
For $S=0.3$, $\rpf=0.032 \approx 0.033 = \rc$.
For $S=3$, the slope of $\sigmac$ exceeds that of $\sigmat$.
The thermosolutal eigenvalues deviate substantially 
from the pure eigenvalues, but  $\sigma_+ \approx \sigmac$ 
near the bifurcation point.
Note change of scale between upper three and lower diagrams.
\newline
}
\label{fig:linplus}
\end{figure}

From figures \ref{fig:linminus} and \ref{fig:linplus},
it can be seen that the thermosolutal thresholds and eigenvalues
are related to the pure thermal and solutal thresholds and eigenvalues.
Such a resemblance is also clearly visible in a numerical study of 
Marangoni (surface-tension-driven) convection with Soret effect \cite{Bergeon}.
This is the relationship we wish to explore in this section.

The simplest classification is by proximity:
a real eigenvalue is primarily thermal if it is closer to the 
pure thermal eigenvalue $\sigmat$ than to the pure solutal eigenvalue
$\sigmac$, i.e. if:
\begin{equation}
\vert \sigma - \sigmat \vert < \vert \sigma - \sigmac \vert
\label{proximity}
\end{equation}
Another possible classification is based on the eigenvector,
specifically on the magnitude of $SC/T$ --
the ratio of the solutal to the thermal contribution in 
the buoyancy force in (\ref{tslphi}).
These two criteria are equivalent.
Indeed, the eigenvalue equation (\ref{lin2Dsys}) 
states that the eigenvectors satisfy:
\begin{subequations}\begin{eqnarray}
\sigma - \sigmat &=& \frac{SC}{T} r \\
\sigma - \sigmac &=& \frac{T}{C} r
\end{eqnarray}\end{subequations}
Thus (\ref{proximity}) becomes:
\begin{equation}
\left| \frac{SC}{T} \right| < \sqrt{ \left| S \right|}
\end{equation}
Thus, an eigenvalue is thermal (solutal) if the corresponding
eigenvector satisfies:
$\left| SC/T \right| <(>) \sqrt{ \left| S \right|}$.

For positive $S$, $\rint$ separates thermal from solutal
portions of the eigenvalue curve:
For $0<S<1$, $\sigma_+$ is thermal and $\sigma_-$ is solutal
for the range $r > \rint$ and vice versa $r < \rint$.
For $S>1$, the opposite holds.
For negative $S$, $\rint$ is the midpoint of the interval
of complex eigenvalues.
Complex eigenvalues, whose real part is equidistant between
the pure thermal and solutal eigenvalues, cannot be classified in this way.
Instead, it is $r_\pm$ which serve as boundaries:
for negative $S$, we classify the real eigenvalues 
$\sigma_+$ as thermal and $\sigma_-$ as solutal 
for the range $r > r_{+}$ and vice versa for $r < r_{-}$.
In figures \ref{fig:linminus} and \ref{fig:linplus}, we can discern
various regimes in which the thermosolutal eigenvalues $\sigma_\pm$ 
adhere closely to the pure solutal and thermal eigenvalues $\sigmat$
and  $\sigmac$. For $|S| \gg 1$, for example, 
$\sigma_+ \approx \sigmac$ for $r$ sufficiently small.
For $|S| \ll L$, $\sigma_\pm$ resemble $\sigmat$ and $\sigmac$
over most of the range shown surrounding $\rint$.
Detailed estimates, omitted here, justify these visual impressions.

Similarity between thermosolutal and pure thermal or solutal eigenvalues 
implies similarity between the thresholds of these problems.
The threshold of the steady bifurcation is $\rpf = L/(L+S)$.
If $|S| \ll L$, then $\rpf \approx 1 = \rt$,
i.e. the thermosolutal threshold approaches the pure thermal threshold.
Conversely, if $|S| \gg L$, then $\rpf \approx L/S = \rc$,
so that the thermosolutal threshold approaches the pure solutal threshold.

More precisely, we can calculate the difference between the thresholds:
\begin{subequations}\begin{eqnarray}
|\rpf-\rt| &=& \left|\frac{L}{L+S} - 1\right| 
             = \left|\frac{L-(L+S)}{L+S}\right| = \left|\frac{S}{L+S}\right|\\
|\rpf-\rc| &=& \left|\frac{L}{L+S} - \frac{L}{S}\right| 
= \left|\frac{LS-L(L+S)}{S(L+S)}\right| = \left|\frac{L^2}{S(L+S)}\right|
\end{eqnarray}\end{subequations}
Taking into account the different sign possibilities for $S$ and $L+S$,
we find that, for $\epsilon \ll 1$,
\begin{subequations}\begin{eqnarray}
|\rpf-\rt| < \epsilon &{\rm ~~if~~}& |S|<L\epsilon \label{rpfrt}\\
|\rpf-\rc| < \epsilon &{\rm ~~if~~}& |S|>\frac{L}{\sqrt{\epsilon}}\label{rpfrc}
\end{eqnarray}\end{subequations}
Indeed, on the left diagrams of figures \ref{fig:linminus} and 
\ref{fig:linplus}, we see that the thermosolutal threshold $\rpf$
(heavy solid curve) is well approximated by the pure thermal threshold
$\rt$ (straight thin dashed curve) for $|S|\lesssim 0.001$
and by the pure solutal threshold $\rc$ (curved thin dashed curve) for 
$|S|\gtrsim 0.03$.
This is precisely the estimate obtained from (\ref{rpfrt}-\ref{rpfrc})
with $\epsilon=0.1$ and $L=0.01$.
The domain $|S| \lesssim 0.1 L$ is analogous to what was termed
the thermal-dominated regime by Bergeon et al. \cite{Bergeon}
while the $|S| \gtrsim 3 L$ is analogous to what was termed
the solutal-dominated regime.
We note further that the eigenvector at the steady bifurcation 
satisfies
\begin{equation}
\frac{SC}{T} = \frac{S}{L}
\end{equation}
Thus, at the bifurcation, either the solutal contribution $SC$ 
or the thermal contribution $T$ dominates the buyoancy force,
according on whether one is in the solutal-dominated or
thermal-dominated regime.

\begin{figure}
\vspace*{-4cm}
\centerline{
\psfig{file=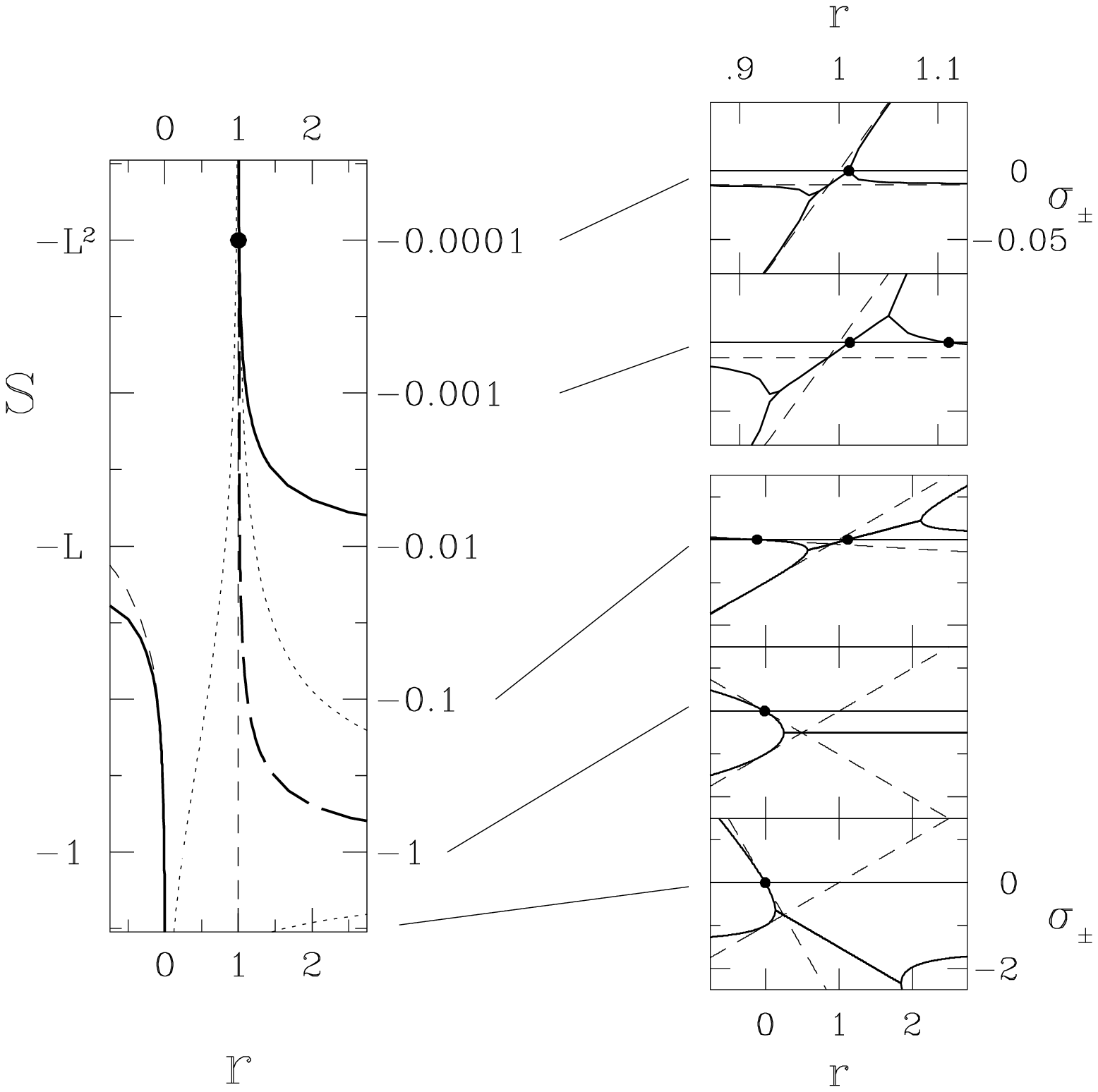,width=20cm}}
\end{figure}
\newpage
\begin{figure}
\caption{{}
Behavior of thermosolutal eigenvalues for negative $S$.\newline
Left: Thresholds for negative values of $S$ plotted on a logarithmic scale.
Solid curves show the thresholds $\rpf=L/(L+S)$ of steady bifurcations. 
Heavy long-dashed curve indicates the thresholds $\rh = (1+L)/(1+S)$ of 
Hopf bifurcations. This curve
appears from $r=\infty$ at $S=-1$ and disappears by meeting the
steady bifurcation curve in a codimension-two point (Bogdanov bifurcation)
at $S_* = -L^2$, $r_*=1/(1-L)$, indicated by a heavy dot.
Between the two dotted curves $r=r_{\pm}$, the eigenvalues are complex;
the right boundary $r=r_{+}$ goes to infinity at $S=-1$.
Light dashed curves show the thresholds $\rt=1$ and $\rc=L/S$ of the 
pure thermal and solutal problems, which are within 0.1 of $\rpf$ for
$S\gtrsim -0.1 L = 0.001$ and $S\lesssim -L/\sqrt{0.1} = -0.03$,
respectively. \newline
Right: Real part $\sigma_\pm$ of the eigenvalues of the thermosolutal
problem as a function of $r$ for representative negative values of $S$.
Straight segments show the real part of complex conjugate 
pairs of eigenvalues. Dots represent bifurcations.
Dashed lines are the eigenvalues $\sigmat$ and $\sigmac$ of the pure 
thermal and solutal problems, with slopes 1 and $S$, respectively
and intersection point $\rint$. 
For $S=-3$, representing $S<-1$, a steady bifurcation with 
$\sigma_+=0$ occurs at $\rpf=-0.00334 \approx -0.00333 = \rc$.
Complex eigenvalues all have negative real part.
For the limiting case $S=-1$, the range of complex eigenvalues extends 
from $r_-=(1-L)/4$ to $r_+=\infty$. All have $\sigma=f=-(1+L)$, with 
$\omega = \sqrt{-g}\rightarrow \infty$ as $r\rightarrow\infty$.
For $S=-0.1$, representing $-1 < S < -L$, 
there is both a Hopf bifurcation at $\rh=1.12$ and a steady 
bifurcation at $\rpf=-0.11$.
For the limiting case $S=-L=-0.01$ (not shown), the pitchfork bifurcation
has disappeared to $+\infty$.
For $S=-0.001$, representing $-L < S < -L^2$, the pitchfork bifurcation 
has reappeared from $-\infty$, with $\sigma_-=0$ at $\rpf=1.11$.
The limiting case $S=-L^2=-0.0001$ is the codimension-two point.
Note change of scale between upper two and lower three diagrams.\newline
----------------------------------------------------------------------------------------------------------------------------
}
\label{fig:linminus}
\end{figure}

\section{Nonlinear analysis}\label{Nonlinear analysis}
\subsection{Derivation of nonlinear model}
We return to the full nonlinear thermosolutal problem 
(\ref{thermosolutal_scaled}) and summarize the derivation 
of a minimal set of amplitude equations.
This model was first introduced by Veronis \cite{Veronis65}
for the thermosolutal problem and later adapted by Platten and
Chavepeyer \cite{PlattenChavepeyer75} for the Soret problem.

The simple spatial dependence (\ref{solform}) is not preserved by the 
nonlinear terms in the governing equations (\ref{thermosolutal_scaled}).
Substituting (\ref{solform}a) and 
\begin{equation}
\hat{\phi}(x,z,t) =\phi(t) \sin(kx) \sin(\pi z)
\label{solformphi}
\end{equation}
where $w=k\phi$, into (\ref{tsfT_scaled}) yields:
\begin{eqnarray}
\ydot (\grad \hat{\phi} \times \grad \hat{T}) 
&=& \dz \hat{\phi} \dx \hat{T} 
- \dx \hat{\phi} \dz \hat{T} \nonumber\\
&=& -\pi k \phi T (\sin^2 (kx) + \cos^2 (kx)) \sin (\pi z) \cos(\pi z) \nonumber\\
&=& -{{\pi} \over 2} w T \sin (2 \pi z) \label{nonlinear_generation}
\end{eqnarray}
and similarly for $\hat{C}$. On the other hand, the dependence
(\ref{solformphi}) is preserved by the nonlinear terms of (\ref{tsfphi_scaled}):
\begin{equation}
\geom^{-2}\ydot (\grad \hat{\phi} \times \grad \lap \hat{\phi}) =
-\ydot (\grad \hat{\phi} \times \grad \hat{\phi}) = 0
\label{preserve}\end{equation}
Expansion (\ref{solform}) is therefore generalized to include terms
of the type (\ref{nonlinear_generation}):
\begin{subequations}
\label{solform_full}
\begin{eqnarray}
\hat{T} &=& 
T (t) \cos (kx) \sin (\pi z) + T_2 (t)\sin (2\pi z) 
\label{sfT}\\
\hat{C} &=& 
C (t)\cos (kx) \sin (\pi z) + C_2 (t)\sin (2\pi z) \label{sfC}
\end{eqnarray}
\end{subequations}
The nonlinear term of (\ref{tsfT_scaled}) on $\hat{\phi}$ and $\hat{T_2}$ yields
\begin{eqnarray}
\ydot (\grad \hat{\phi} \times \grad \hat{T_2}) 
= \dz \hat{\phi} \dx \hat{T_2} - \dx \hat{\phi} \dz \hat{T_2} \nonumber\\
= 0 - 2\pi w T_2 \cos (kx) \sin (\pi z) \cos (2\pi z) \nonumber \\
= \pi w T_2 \cos (kx) (\sin (\pi z) - \sin (3\pi z))
\end{eqnarray}
The expansion (\ref{sfT}-\ref{sfC}) is truncated by neglecting the term 
$\cos (kx) \sin (3\pi z)$.
This procedure is quantitatively correct for small amplitudes and, 
as is often the case, qualitatively accurate even for moderate amplitudes. 
The validity and limitations of this truncation are discussed in
\cite{Knobloch80,KnoblochProctor,KnoblochMooreToomreWeiss,Rucklidge,KnoblochProctorWeiss}.

Substituting (\ref{sfT}-\ref{sfC}), (\ref{solformphi}), and (\ref{solform}c)
into (\ref{thermosolutal_full}) yields:
\begin{equation}
\frac{d}{dt}
\left(\begin{array}{c} T \\ C \\ w \\ T_2 \\ C_2 \end{array} \right) = 
\left(\begin{array}{ccccc} -1 & 0 & 1 & 0 & 0 \\
                           0 & -L & 1 & 0 & 0 \\
                           Pr & PSr 
                                                   & -P & 0 & 0 \\
                           0 & 0 & 0 & -4\pi^2/\geom^2 & 0 \\
                           0 & 0 & 0 & 0 & -4\pi^2 L/\geom^2 \\ \end{array} \right)
\left(\begin{array}{c} T \\ C \\ w \\ T_2 \\ C_2 \end{array} \right) 
+\pi w \left(\begin{array}{c} T_2 \\ 
                                                 C_2 \\
                                                    0 \\
                                              -T /2 \\
                                              -C /2 \\ \end{array} \right)
\label{full5Dmodel}\end{equation}

In the linear part of (\ref{full5Dmodel}), $(T,C,w)$ is decoupled
from $(T_2,C_2)$, and the eigenvalues of the latter system are
always negative. Hence, the linear stability analysis of (\ref{full5Dmodel})
is that already carried out in section \ref{Linear analysis}, 
assuming again that $P \rightarrow \infty$.

The remainder of section \ref{Nonlinear analysis} is devoted to analyzing 
the exact steady states of (\ref{full5Dmodel}).
Surprisingly, we will find that calculating these steady states 
reduces to calculating the eigenvalues of a $2 \times 2$ matrix
closely related to the linear stability matrix (\ref{lin2Dsys}).

\subsection{Steady states of the nonlinear model}
Although (\ref{full5Dmodel}) is a five-dimensional nonlinear system,
its form allows its steady solutions to be calculated analytically
\cite{Veronis65}.
Steady solutions of (\ref{full5Dmodel}) satisfy:
\begin{eqnarray}
\nonumber\label{steady5D}
\end{eqnarray}
\vspace*{-.5cm}
\addtocounter{equation}{1}
\begin{displaymath}
\left(\begin{array}{c} 0 \\ 0 \\ 0 \\ 0 \\ 0 \end{array} \right) = 
\left(\begin{array}{ccccc} -1 & 0 & 1 & 0 & 0 \\
                           0 & -L & 1 & 0 & 0 \\
                           Pr & PSr 
                                                   & -P & 0 & 0 \\
                           0 & 0 & 0 & -4\pi^2/\geom^2 & 0 \\
                           0 & 0 & 0 & 0 & -4\pi^2 L/\geom^2 \\ \end{array} \right)
\left(\begin{array}{c} T \\ C \\ w \\ T_2 \\ C_2 \end{array} \right) 
+\pi w \left(\begin{array}{c} T_2 \\ 
                              C_2 \\
                                0 \\
                          -T /2 \\
                          -C /2 \\ \end{array} \right)\;\;\;\;\;\;
\begin{array}{c} (\ref{steady5D}{\rm a}) \\ (\ref{steady5D}{\rm b}) \\ 
                 (\ref{steady5D}{\rm c}) \\ (\ref{steady5D}{\rm d}) \\ 
                 (\ref{steady5D}{\rm e}) \end{array}
\end{displaymath}
Note that (\ref{steady5D}c), which was derived as (\ref{velslave})
from the time-evolution equation under the assumption of large $P$, 
is here merely a consequence of the search for steady states
and of (\ref{preserve}).
Using (\ref{steady5D}c), (\ref{steady5D}d), and (\ref{steady5D}e) 
to eliminate $w$, $T_2$, and $C_2$ yields:
\begin{equation}
\left(\begin{array}{cc} r-1 & Sr \\ r & Sr-L \end{array}\right) 
\left(\begin{array}{c} T \\ C \end{array}\right) 
=\frac{\geom^2 w^2 }{8}
\left(\begin{array}{cc} 1 & 0 \\ 0 & {1\over L} \end{array}\right)
\left(\begin{array}{c} T \\ C \end{array}\right)
\label{nonlinmatrix}\end{equation}

Equation (\ref{nonlinmatrix}) is of the form of an eigenvalue problem:
\begin{equation}
\left(\begin{array}{cc} r-1 & Sr \\ Lr & L(Sr-L) \end{array}\right) 
\left(\begin{array}{c} T \\ C \end{array}\right) =
E\left(\begin{array}{c} T \\ C \end{array}\right)
\label{non2Dsys}\end{equation}
with 
\begin{equation}
E \equiv \frac{\geom^2 w^2}{8} = 
\frac{\geom^2 r^2}{8} (T+SC)^2
\label{Edef}\end{equation}
playing the role of an eigenvalue.
After the eigenvalues $E$ of (\ref{non2Dsys}) are found,
the relative amplitude of components $T$ and $C$ is
given by the eigenvectors.
Note that the original five-dimensional steady-state problem
(\ref{steady5D}) is also of this type. Then, 
$-\pi w$ plays the role of a generalized eigenvalue
and the generalized eigenvectors give the relative proportions of
the five components.

A related version of this reduction has been shown \cite{HollingerLucke98}
to be valid for the full PDEs governing binary
fluid convection with Soret effect with realistic boundary conditions,
and has been used \cite{HollingerLuckeMuller} to derive a
sophisticated and realistic 11-mode model.
Hollinger et al. \cite{HollingerLucke98,HollingerLuckeMuller} 
demonstrate that the velocity field can be very well approximated 
by a single spatial mode and can be adiabatically eliminated. 
The nonlinearities are then only those which advect the
temperature and concentration, and contain only the amplitude $w$ of the
vertical velocity. The temperature and concentration amplitudes
are then solutions to a linear system depending on $w$.
Hollinger et al.'s work provides evidence that our reformulation of 
(\ref{steady5D}) as an eigenvalue problem, while strictly valid
only for the minimal five-mode Veronis model of thermosolutal 
convection, is a manifestation of a quite general property of 
double-diffusive convection.

$E$ can plausibly be called the energy of the convective state,
since the kinetic energy density is:
\begin{eqnarray}
{\mathcal E} &=& \frac{k}{2\pi}
\int_0^{2\pi/k} dx \int_0^1 dz \;\frac{1}{2} (\hat{u}^2+\hat{w}^2) \nonumber\\
  &=& \frac{k}{2\pi} \int_0^{2\pi/k} dx \int_0^1 dz \frac{1}{2} 
\left(\left(\frac{\pi w}{k}\sin (kx) \cos(\pi z)\right)^2 + (w\cos (kx) \sin(\pi z))^2\right) \nonumber\\
  &=& \frac{\geom^2 w^2}{8 k^2}
   = \frac{1}{k^2} E 
\end{eqnarray}
The most common experimentally measured quantity is the
convective heat transport ${\mathcal N}-1$, which is related to $E$ via:
\begin{eqnarray}
{\mathcal N} - 1 &=& \int_0^{2\pi/k} dx \int_0^1 dz \;\hat{w}\;\hat{T} 
  = \int_0^{2\pi/k} dx \int_0^1 dz \;w\;T\; \cos^2(kx)\sin^2(\pi z) \nonumber\\
  &=& \frac{\pi}{2k} \;w\;T = \frac{\pi}{2k} \frac{w^2}{E+1} 
= \frac{4\pi}{k q^2} \frac{E}{E+1} \label{nusselt}
\end{eqnarray}
where we have used $w-T=ET$ from (\ref{non2Dsys}).

The matrix:
\begin{eqnarray}
\left(\begin{array}{cc} r-1 & Sr \\ Lr & L(Sr-L) \end{array}\right) 
&=& \left(\begin{array}{cc} -1 & 0 \\ 0 & -L^2 \end{array}\right) 
+r\left(\begin{array}{cc} 1 & S \\ L & LS \end{array}\right)\label{nonmatrix}\\
\tilde{M} &=&  \tilde{M_0} + r \tilde{M_1} \nonumber
\end{eqnarray}
is very similar to the  matrix (\ref{linmatrix}). 
$\tilde{M}_0$ again describes diffusion, and $\tilde{M}_1$
now describes the combined effects of advection and nonlinear saturation.
The interpretation of the behavior of the eigenvalues of (\ref{nonmatrix})
is, of course, different from that of (\ref{linmatrix}).
In the linear system, negative or complex eigenvalues characterize 
infinitesmal perturbations which decay and/or oscillate.
For the nonlinear problem, negative or complex values of $E$ 
are forbidden by virtue of definition (\ref{Edef}), and imply 
non-existence of steady solutions for certain ranges of $r, S, L$.
Since only real positive values of $E$ are meaningful,
we will not introduce separate notation for real and
imaginary parts of $E$. For $E$ real and positive, 
we define $A=\pm\sqrt{E}$ as the convection amplitude.

\subsection{Bifurcations}

Most of the results of section \ref{Linear analysis} concerning
the eigenvalues $\sigma_\pm$ are easily 
modified to apply to the energy $E$ of the nonlinear 
steady states, merely by substituting
\begin{subequations}\begin{eqnarray}
L \rightarrow L^2  \\
S \rightarrow LS
\end{eqnarray}\label{substitute}\end{subequations}
When we use the same notation for the nonlinear and linear problems,
we add tildes to designate nonlinear quantities.

\begin{figure}
\vspace*{-3.4cm}
\centerline{
\psfig{file=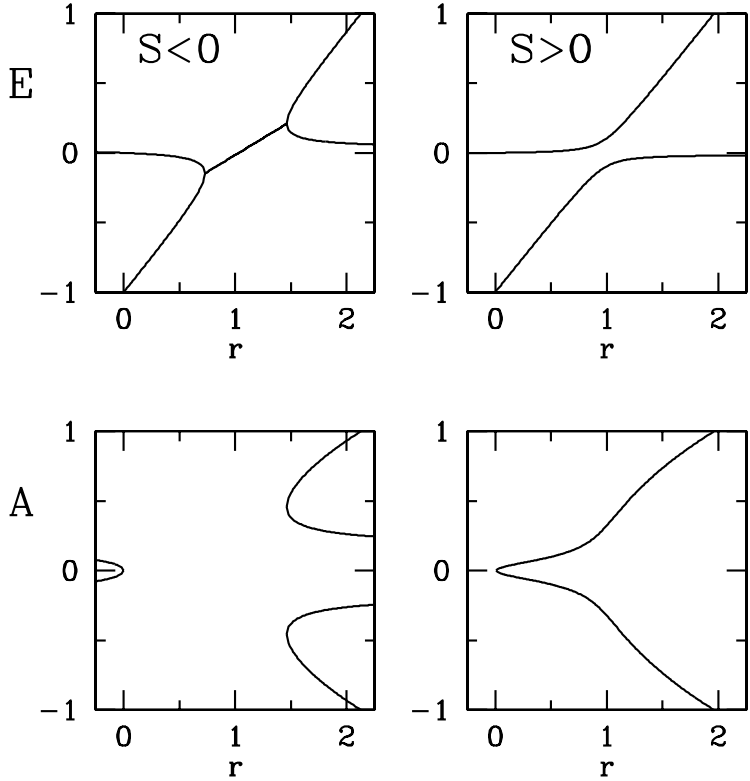}}
\vspace*{-8cm}
\caption{{}
$E$ and $A=\pm\sqrt{E}$ (for $E$ real and positive) for $S=-3$ and $S=1$.
For $S=-3$, $E$ undergoes complex coalescence.
The point at which $E$ becomes complex corresponds
to a pair of saddle-node bifurcations in $A$.
For $S=1$, $E$ undergoes avoided crossing.
This leads to a distinction in $A$ between 
low-amplitude Soret and high-amplitude Rayleigh regimes.
Pitchfork bifurcations are at $r=0.0099$ for $S=1$
and at $r=-.003344$ for $S=-3$.}
\label{fig:squareroot}
\end{figure}

Just as we did for the linear problem,
we can define nonlinear pure thermal and solutal solutions 
by setting the coupling terms in (\ref{non2Dsys}) to zero:
\begin{subequations}\begin{eqnarray}
E_T = r-1 = r-\rt \\
E_C = L(Sr - L) = LS(r-\rc)
\end{eqnarray}
Again, the coupling term 
\begin{equation}
\tilde{\coupa}\tilde{\coupb} = LSr^2
\end{equation}\end{subequations}
is proportional to, and has the same sign as, $S$.
Hence the eigenvalues $E$ undergo avoided crossing near 
the intersection point of the two pure solutions
\begin{equation}
\rintnl = \frac{1-L^2}{1-LS}
\end{equation}
if $S$ is positive, and complex coalescence if $S$ is negative.
These possibilities are illustrated in figure \ref{fig:squareroot},
along with the resulting consequences for $A=\pm\sqrt{E}$.

The solutions to the eigenvalue problem (\ref{non2Dsys}) are:
\begin{subequations}
\begin{eqnarray}
E&=& {{E_T+E_C}\over 2} \pm 
\sqrt{ \left({{E_T-E_C}\over 2}\right)^2 + SLr^2} \\
    &=& {1\over 2} \left[(1+LS)r - (1+L^2) \right] \nonumber\\
&&\;\;\;\;\ \pm {1\over 2}\sqrt{(1 + LS)^2 r^2 - 2(1-L^2)(1-LS)r + (1 - L^2)^2}\\
    &\equiv& \tilde{f}(r) \pm \sqrt{\tilde{g}(r)}
\end{eqnarray}
\label{2Damps}\end{subequations}
This expression for $E$ is quite similar to (\ref{2Deigs}) for the eigenvalues.
Detailed results for representative values of $S$ are shown in the 
two large figures \ref{fig:nonminus} and \ref{fig:nonplus}.

Just as $\lambda=0$, $(T,C)=(0,0)$ is a solution of (\ref{lin2Dsys}), 
it is also true that $E=0$, $(T,C)=(0,0)$ is always a solution to (\ref{non2Dsys}).
The existence of the solution $E=0$ for all $r$ reflects the fact
that the conductive profile in a motionless fluid remains a solution
for all Rayleigh numbers, although not necessarily a stable one.
Equations (\ref{2Deigs}) and (\ref{2Damps}) give the nontrivial solutions 
for $\lambda$ and $E$, which are zero only at isolated values of $r$.

The values (\ref{2Damps}) for $E$ real form hyperbolas:
\begin{equation}
\left(E + \frac{LS+L^2}{1+LS} - (1+LS)\left(r - \frac{(1-L^2)(1-LS)}{(1+LS)^2}\right)\right) 
\left(E + \frac{LS+L^2}{1+LS}\right) = SL\;\frac{(1-L^2)^2}{(1+LS)^2} \equiv \tilde{\Sgen}
\label{hyperbolanon}\end{equation}
in the $(r,E)$ plane.
(In the exceptional case $S=-1/L$, the solutions form a parabola.)
Just as we found that $r$ was a single-valued function (\ref{rfcnl}) of 
$\sigma$, here $r$ is also a single-valued function of $E$.
(This is again because the zero determinant of matrix $\tilde{M_1}$ in 
(\ref{nonmatrix}) leads to a horizontal asymptote for the hyperbola.)
We use the substitutions (\ref{substitute}) to transform $r(\sigma)$ into 
$r(E)$:
\begin{equation}
r = \frac {\sigma_{\pm}^2 + \sigma_{\pm}(1+L) + L} {\sigma_{\pm}(1+S) + (S+L)} 
\;\;\longrightarrow\;\;
r = \frac {E^2 + E(1+L^2) + L^2} {E(1+LS) + L(S+L)} 
\label{rfcna}\end{equation}
a result which can verified by inverting (\ref{2Damps}).
Similar formulas are found in, e.g. \cite{Veronis65,Sani,PlattenLegros}.
Thus each value of $E$ except $-L(S+L)/(1+LS)$ is achieved exactly once,
a feature mentioned in section \ref{Linear analysis} 
for the growth rates $\sigma$, but more 
important in the context of amplitudes of steady states;
see figures \ref{fig:nonminus} and \ref{fig:nonplus}.
It is remarkable that this property continues to hold even for the 
far more complicated Soret problem with rigid boundaries and adequate 
spatial resolution \cite{HollingerLuckeMuller}.
In \cite{HollingerLuckeMuller}, it is also found that 
$r$ is a single-valued function of the energy, as measured by 
the square of the vertical velocity amplitude.
This function, like (\ref{rfcna}), is the ratio of a quadratic 
to a linear function of $E$, although with
coefficients more complicated than those of (\ref{rfcna}).

Steady bifurcations from the conductive profile occur at values of 
$r$ at which $E=0$.
Equation (\ref{rpf}) gives the values $\rpf$ at which $\sigma = 0$.
The substitutions (\ref{substitute}) leave expression (\ref{rpf}) unchanged:
\begin{equation}
\rpf = \frac{L}{S+L}
\;\;\longrightarrow\;\;
\tilde{r}_{\rm PF} = \frac{L^2}{LS+L^2} = \frac{L}{S+L} = \rpf
\end{equation}
This result can be verified by setting $E=0$ in (\ref{2Damps}) or 
(\ref{rfcna}).
The fact that the ``linear'' ($\sigma=0$) and ``nonlinear'' 
($E=0$) values of $\rpf$ are identical reflects the fundamental 
bifurcation-theoretic fact
that a change of sign of a real eigenvalue signals the transverse
intersection of two or more steady branches, i.e. a steady bifurcation.
The fact that $E=A^2$ (rather than $A$) changes sign at $\rpf$ 
identifies these as pitchfork bifurcations, since either sign of $A$
is permitted.
Equations (\ref{nusselt}) and (\ref{rfcna}) can be used to calculate
the slope of the convective heat transport at the bifurcation:
\begin{equation}
\left.\frac{{\rm d}({\mathcal N} - 1)}{{\rm d} r} \right|_{r=\rpf}
= \left.\frac{4\pi}{kq^2} \;\frac{1}{(E+1)^2} 
\;\left(\frac{{\rm d} r}{{\rm d} E}\right)^{-1} \right|_{E=0}
= \frac{4\pi}{kq^2} \;\frac{L(S+L)^2}{S+L^3}
\label{nussbif}\end{equation}
very similar to the analogous formula derived for the Soret problem in \cite{Brand}.

If $S<0$, complex solutions to (\ref{2Damps}) are possible.
The interval of $r$-values over which this is so, delimited by $\tilde{g}=0$,
is easily obtained 
using substitutions (\ref{substitute}) in (\ref{complex_range_lin}):
\begin{equation}
\frac{1-L^2}{(1+\sqrt{-LS})^2}
\equiv \tilde{r}_- < r < \tilde{r}_+ \equiv
\frac{1-L^2}{(1-\sqrt{-LS})^2}
\label{complex_range_non}\end{equation}
For the linear problem, the endpoints of the complex interval
(\ref{complex_range_lin}) for $\lambda$ divided oscillatory
evolution from monotonically growing (if $\sigma > 0$) or decaying 
(if $\sigma < 0$) evolution.
For the nonlinear problem, the endpoints of the complex interval
(\ref{complex_range_non}) mark the appearance or disappearance
of real solutions $A$; see figure \ref{fig:squareroot} for $S<0$ and
figure \ref{fig:nonminus} for $S=-10$.
These are saddle-node bifurcations: the simultaneous
creation of four branches of steady solutions $A$, corresponding
to two different values of $E=A^2$.
Since $E$ must be non-negative as well as real, the occurrence
of saddle-node bifurcations at $\tilde{r}_-$ or $\tilde{r}_+$ 
requires that ${\mathcal R}(E) = \tilde{f}\geq 0$.
Substituting (\ref{complex_range_non}) into
\begin{equation}
\tilde{f} (\tilde{r}_\pm) = 
\frac{1}{2} \left[\tilde{r}_\pm-1 + LS(\tilde{r}_\pm-L^2)\right]
\end{equation}
we calculate that saddle-node bifurcations occur at $\rsn \equiv \tilde{r}_+$
if
\begin{equation}
-{1\over L} < S < -L^3
\label{snrange}\end{equation}

\begin{figure}
\centerline{
\psfig{file=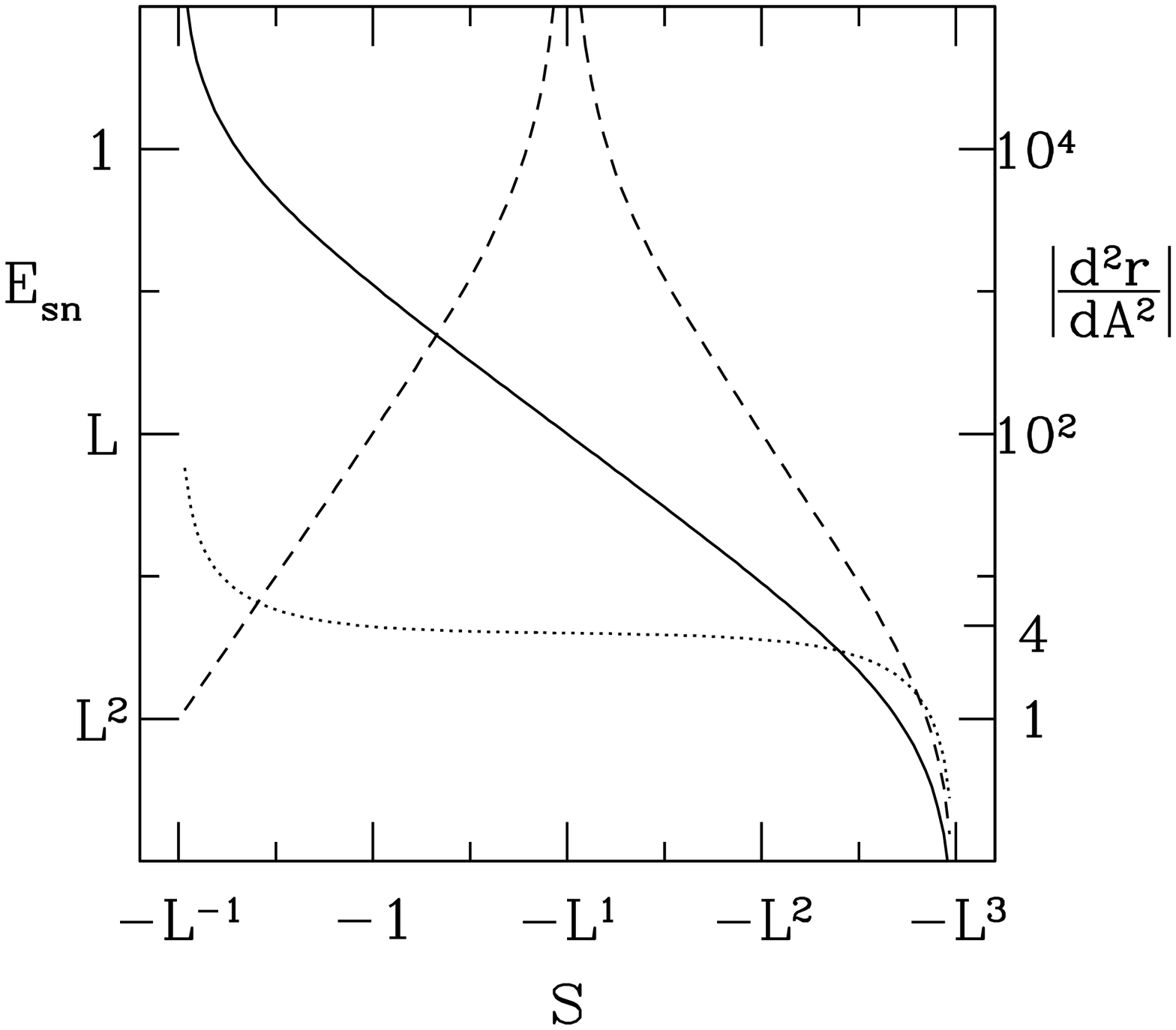,width=8cm}}
\caption{{}
Saddle-node bifurcation for $-1/L < S < -L^3$.
The solid curve shows $\Esn$, the energy at the saddle-node 
bifurcation at $\rsn$.
$\Esn \approx \sqrt{|LS|}$ except very near $-1/L$, 
where it diverges, and near $-L^3$, where it goes to zero.
The dotted curve shows $d^2r/dA^2(\rsn)$, the curvature of the
saddle-node bifurcation, which remains near 4
except near the endpoints $-1/L$ and $-L^3$.
For contrast, the dashed curve shows $d^2r/dA^2(\rpf)$,
the curvature of the pitchfork bifurcation,
which varies greatly everywhere over this interval, especially near $-L$,
where $\rpf$ and $d^2r/dA^2(\rpf)$ approach $\pm\infty$.
}
\label{fig:esn}
\end{figure}

Figure \ref{fig:esn} characterizes the saddle-node bifurcation
over the domain $-1/L < S < -L^3$ of its existence.
Over most of the range, the value $\Esn = E(\rsn)$ 
varies with $S$ like $1/\sqrt{-SL}$ and $\rsn$ varies like
$1+2\sqrt{-SL}$. 
When the saddle-node bifurcation appears at $S=-1/L$, 
we have $\rsn = \Esn = \infty$.
When the saddle-node bifurcation disappears at the degenerate pitchfork
at $S=-L^3$, we have $\rsn=1/(1-L^2)$, $\Esn=0$. 
The curvature $d^2 r/dA^2 (\rsn)$ at the saddle-node bifurcation is
also shown.
$d^2 r/dA^2 (\rsn) \approx 4$ over much of the range.
In contrast, the curvature at the pitchfork bifurcation 
$d^2 r/dA^2 (\rpf) \sim (d ({\mathcal N}-1)/dr )^{-1} (\rpf)$
(see equations (\ref{rfcna}) and (\ref{nussbif}))
varies over many orders of magnitude (even away from its divergence at $S=-L$).
This great difference in curvatures is one of the factors
giving the thermosolutal bifurcation diagrams, such as the $S=-0.003$
case in figure \ref{fig:nonminus}, their characteristic appearance. 

We note that (\ref{snrange}) is the transformation via
substitutions (\ref{substitute}) of the range of existence
(\ref{hopfrange}) of Hopf bifurcations of the linear problem.
Yet the criterion for a Hopf bifurcation, where $\lambda$ has
zero real part ($f = 0$) and finite imaginary part ($g \leq 0$)
has no significance for a nonlinear steady state $E$.
We explain the correspondence between the range of existence
of Hopf and saddle-node bifurcations as follows.

Although a Hopf bifurcation has no analogue for nonlinear steady states,
the codimension-two point of the linear analysis,
at which the pitchfork and Hopf bifurcations coalesce,
does have a relevant nonlinear analogue.
This is because at the codimension-two point,
the pitchfork and Hopf bifurcations necessarily coalesce with a third point:
that at which the eigenvalues become complex.
(See $S=-L^2$ case of figure \ref{fig:linminus}.)
That is, the conditions defining a steady bifurcation, 
$f\pm\sqrt{g}=0$, $g\geq0$ and a Hopf bifurcation, $f=0$, $g\leq0$ 
together imply $f=g=0$.
For the nonlinear analysis, the codimension-two point marks
the coalescence of the pitchfork bifurcation, defined by
$\tilde{f}\pm\sqrt{\tilde{g}}=0$, $\tilde{g}\geq0$,
and the saddle-node bifurcation, defined by 
$\tilde{f}\geq0$, $\tilde{g}=0$. These conditions together
again imply $\tilde{f}=\tilde{g}=0$.

We use (\ref{substitute}) to transform the codimension-two point of the 
linear problem to the codimension-two point of the nonlinear problem.
\begin{eqnarray}
S_* =- L^2 &\longrightarrow& 
L\tilde{S}_* = -L^4 {\rm ~~i.e.~~} \tilde{S}_* = -L^3 \label{codim2non}\\
r_*=\frac{1}{1-L} &\longrightarrow& \tilde{r}_*=\frac{1}{1-L^2} 
\end{eqnarray}
Whereas the linear codimension-two point $S_*=-L^2$ marks the 
limit of existence of the Hopf bifurcation, the nonlinear 
codimension-two point $\tilde{S}_*=-L^3$ marks the limit of existence of 
the saddle-node bifurcations. 
For $S > -L^3$, the pitchfork bifurcation is forwards,
with a pair of nontrivial solutions branching right towards $r > \rpf$.
For $S < -L^3$, the pitchfork bifurcation is backwards,
with nontrivial solutions branching left towards $r < \rpf$.

The correspondence between the lower endpoint
of ranges (\ref{hopfrange}) and (\ref{snrange})
is also easily explained.
As $S \downarrow -1$, the Hopf bifurcation point $\rh$ diverges to $+\infty$,
as seen in figure \ref{fig:linminus}.
This event coincides with the divergence of the right endpoint $r_+$
of the interval of complex eigenvalues $\lambda$, as the set
$(r,\pm\omega(r))$ evolves from an ellipse to a righward-opening parabola.
Similarly, the right endpoint $\tilde{r}_+ = \rsn$ of the interval 
of complex (forbidden) values of $E$ diverges to $+\infty$
as $LS \downarrow -1$ or $S \downarrow -1/L$, as seen in figure
\ref{fig:nonminus}.

The saddle-node bifurcation for the nonlinear thermosolutal and 
the Soret problems is well-known, as are the Hopf bifurcation
and the codimension-two point of the linear problem.
Indeed, these are the basic features that originally inspired the great 
interest evoked by binary fluid convection.
However, the relationship between these phenomena has not been 
previously formulated.
One of the advantages of the idealized free-slip thermosolutal 
model is that the codimension-two points have the simple forms 
$S_* = -L^2$ and $\tilde{S}_* = -L^3$.
However this {\it scaling} is quite general.
For the Soret problem with free-slip permeable boundaries
and finite Prandtl number $P$ 
\cite{HurleJakeman,Brand,Knobloch86}, 
\begin{equation}
S_* = -L^2 \;\;\frac{P+1}{P+L(1+L)(P+1)}
\;\;\;\;\;\;\;\;\;\;\;\;\;\;\;\;\;
\tilde{S}_* = -\frac{L^3}{1+L+L^2+L^3}
\end{equation}
Sch\"opf and Zimmermann \cite{Schopf} observe the scaling
$S_* = -L^2$ and $\tilde{S}_* = -L^3$ in amplitude equations
calculated for the Soret problem with rigid impermeable boundaries.
Using their 11-mode minimal model, Hollinger et al. \cite{HollingerLuckeMuller} 
calculate
\begin{equation}
S_* = -L^2 \;\;
\frac{0.35 \; P + 0.18}{P + 0.28\; L}
\;\;\;\;\;\;\;\;\;\;\;\;\;\;\;\;\;
\tilde{S}_* = -L^3 \;\; \frac{1.96}
{1 + 1.04 \; L + 0.37 \; L^2 + 0.97 \; L^3}
\end{equation}
Exact expressions for the lower bounds for the existence of the 
Hopf ($S > -1$ in our case) and saddle-node ($S > -1/L$ in our case) 
bifurcations are not 
as readily available, but also continue to hold approximately.
This provides empirical evidence that the relationship 
between the complex eigenvalues of the linear growth-rate problem
and the saddle-node bifurcation of the nonlinear steady-state problem
continues to hold, at least approximately, even for a
more complicated and realistic case, and thus that
this relationship is a fundamental feature of binary fluid convection.

\begin{figure}
\vspace*{-4cm}
\centerline{
\psfig{file=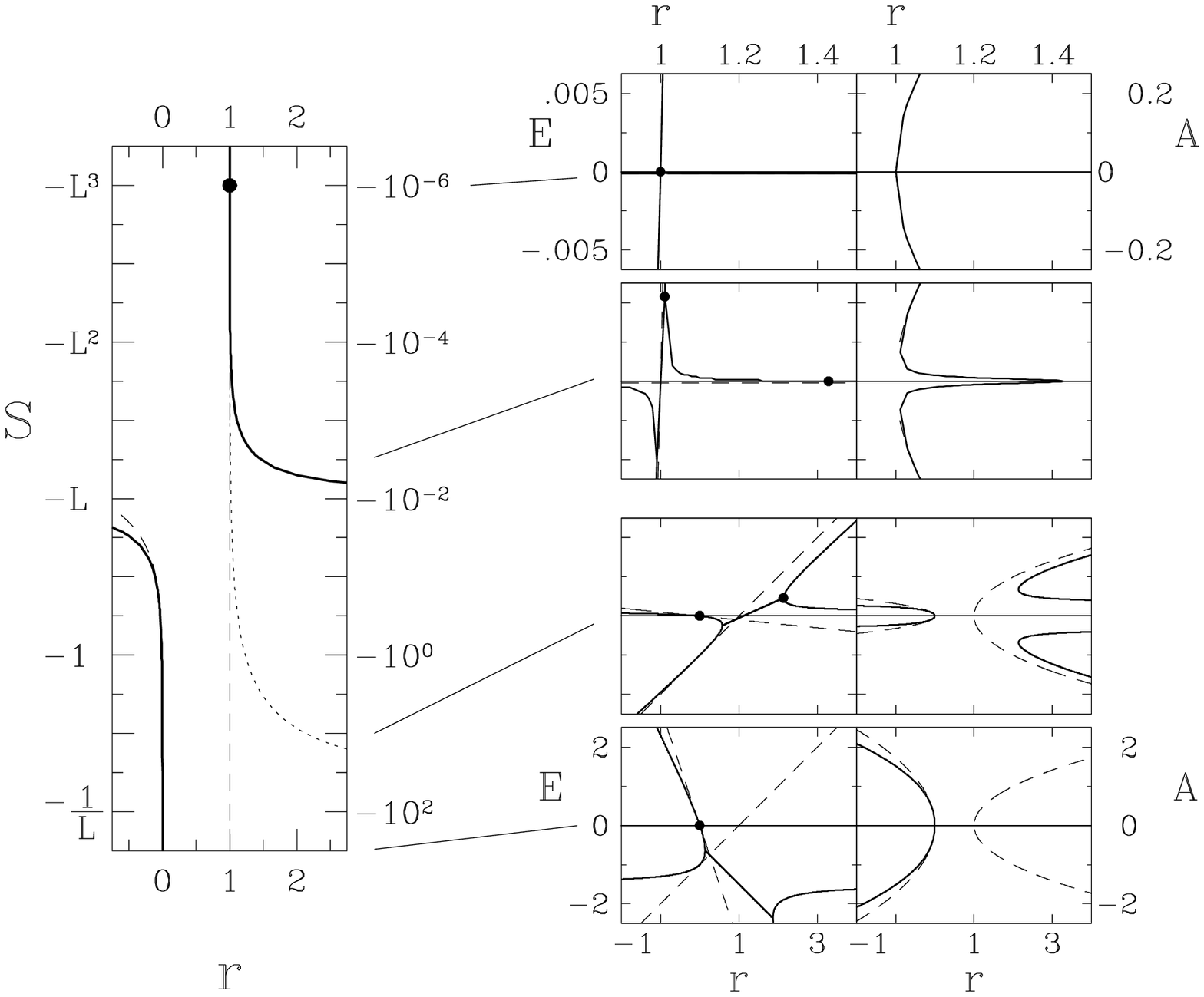,width=20cm}}
\end{figure}
\newpage
\begin{figure}
\caption{{}
Bifurcation diagrams for negative $S$. \newline
Left: thresholds for negative values of $S$ plotted on a logarithmic scale.
Solid curves show the thresholds $\rpf=L/(L+S)$ of steady bifurcations.
Dotted curve shows the thresholds $r_{SN}$ for saddle-node bifurcations.
This curve appears from $r=+\infty$ at $S=-1/L$ and disappears by
meeting the steady bifurcation curve in a codimension-two point
(degenerate pitchfork) 
at $\tilde{S}_*=-L^3$, $\tilde{r}_*=1/(1-L^2)$, indicated by a heavy dot.
Dashed curves show the pitchfork bifurcation thresholds $\rt=1$ 
and $\rc=L/(L+S)$ of the pure thermal and solutal problems.\newline
Right: Energy $E$ and amplitude $A$ of steady nonlinear solutions to 
thermosolutal problem
as a function of $r$ for representative negative values of $S$.
Dashed curves on lower two sets of diagrams show energy and 
amplitude of pure thermal and solutal solutions.
For $S=-300$, representing $S<-1/L$, there is one branch of
real solutions $A$, bifurcating towards negative $r$ and resembling the
pure solutal branch.
At $S=-1/L$ (not shown), the set $(r,E)$ is a parabola and
a pair of saddle-node bifurcations descends from $r=\infty$.
For $S=-10$, representing $-1/L < S < -L$, 
the resulting pairs of disconnected branches can be seen
inside and rather close to the pure thermal branch.
The other solutions, branching towards negative $r$,
are still present. They, and the pure solutal solutions
which they resemble, are of greatly decreased amplitude.
At $S=-L$ (not shown), the pitchfork bifurcation disappears at $r = -\infty$
to reappear at $r = +\infty$. (The pure solutal branch,
increasingly small and distant, continues to exist until $S=0$.)
For $S=-0.003$, representing $-L < S < -L^3$, the pitchfork
bifurcation has descended to $r=1.429$ and connects the two pairs 
of branches arising from the saddle-node bifurcations.
Here, and in the next case,
the thermal branch resembles the upper branch of solutions
too closely to be distinguished from it in the figure,
whereas the solutal branch cannot be distinguished from
the $r$-axis.
At $S=-L^3=-10^{-6}$, the saddle-nodes and pitchfork coalesce
in a codimension-two point separating subcritical from supercritical
pitchfork bifurcations.
Note change of scale between upper two and lower two diagrams.\newline
----------------------------------------------------------------------------------------------------------------------------
}
\label{fig:nonminus}
\end{figure}

\begin{figure}
\vspace*{-4cm}
\centerline{
\psfig{file=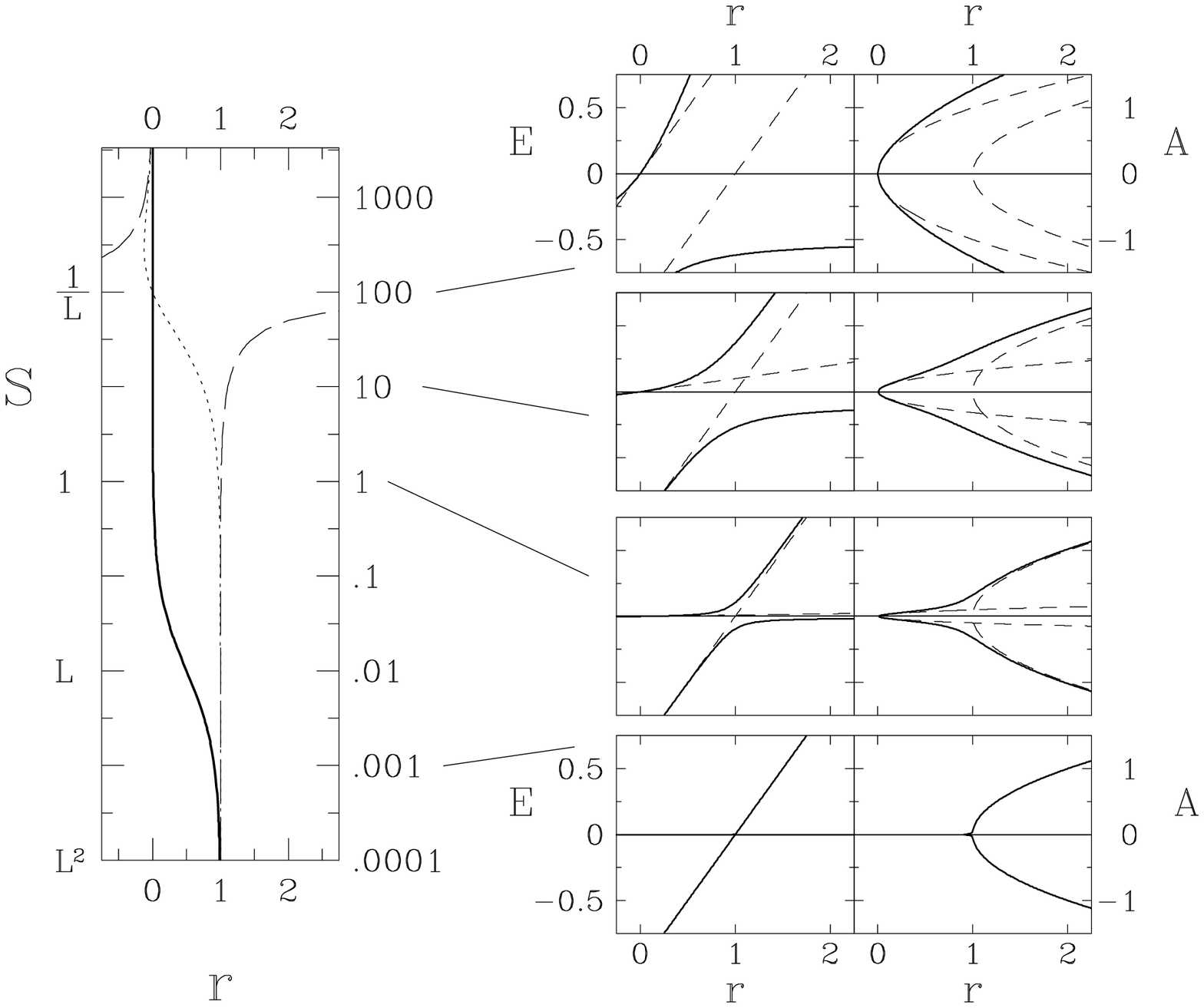,width=20cm}}
\end{figure}
\newpage
\begin{figure}
\vspace*{-2cm}
\caption{{}
Bifurcation diagrams for positive $S$. \newline
Left: Thresholds for positive values of $S$ plotted on a logarithmic scale.
Solid curve show the thresholds $\rpf=L/(L+S)$ of steady bifurcations.
Long-dashed curve shows value $\rintnl$ at which the pure thermal
and solutal branches intersect.
Dotted curve shows value $\rmidnl$ at which the asymptotes of the
hyperbola $E$ intersect, and at which the slope of $E$ and the
curvature of $A$ change most rapidly.\newline
Right: Energy $E$ and amplitude $A$ of steady nonlinear solutions to 
thermosolutal problem
as a function of $r$ for representative positive values of $S$.
Dashed curves show energy and amplitude of pure thermal 
and solutal solutions.
At $S=0.001$, representing $L^3<S<L$ , the thermosolutal branch hugs
the thermal branch.
At $S=1$ and $S=10$, representing $L<S<1/L$, the pure solutal branch 
has descended below the pure thermal branch, ``pulling'' the thermosolutal 
branch with it. 
At $S=100$, representing $S\geq 1/L$, the solutal branch lies above 
the thermal branch. The thermosolutal branch, initially tangent
to the solutal branch, lies above both pure branches.
All bifurcation diagrams use the same scale.\newline
}
\label{fig:nonplus}
\end{figure}

\subsection{Soret and Rayleigh regimes}

In section \ref{Linear thermal and solutal regimes}, 
we saw that the eigenvalues could be characterized as 
primarily thermal or primarily solutal according to their distance
from the pure thermal or solutal eigenvalues.
Here we will discuss various ways of classifying 
the nonlinear solution branches in this way.
This classification is more significant since we are interested in 
entire branches of nonlinear steady states, whereas eigenvalues are 
of interest primarily at the thresholds.

We can apply the classification by proximity analogous to
that we used for the eigenvalues:
a nonlinear steady state is primarily thermal
if it is closer to the pure thermal than to the solutal branch, i.e. if
\begin{equation}
\vert E - E_T \vert < \vert E - E_C \vert
\label{proximitynon}\end{equation}
and primarily solutal otherwise.
Just as we did for the eigenvalues and eigenvectors, we
can show that this criterion is equivalent to one based
on the magnitude of $SC/T$, which is
the ratio of the solutal to the thermal contribution in
the definition of the convective amplitude $A \propto T+SC$
as well as in the buoyancy force.
The eigenvalue equation (\ref{non2Dsys}) 
states that the nonlinear steady states satisfy:
\begin{subequations}\begin{eqnarray}
E - E_T &=& \frac{SC}{T} r \\
E - E_C &=& \frac{LT}{C} r
\end{eqnarray}\end{subequations}
Thus (\ref{proximitynon}) becomes:
\begin{equation}
\left| \frac{SC}{T} \right| < \sqrt{ \left| LS \right|}
\end{equation}
and so a nonlinear steady state is thermal (solutal)
if $SC/T <(>) \sqrt{LS}$.

We first consider negative $S$.
Four ranges of $S$ can be distinguished,
as can be seen on figure \ref{fig:nonminus}.
For $-L^3 < S < 0$, a pair of branches bifurcates towards positive $r$;
these nonlinear steady states are all thermal.
For $-L < S < -L^3$, the pitchfork bifurcation is backwards
and both $E_+$ and $E_-$ are real and positive for $\rsn < r < \rpf$. 
Thermal and solutal branches are separated by $\tilde{r}_+ = \rsn$,
where $SC/T = \sqrt{LS}$. The lower branch $E_-$ is solutal
(but there exist no corresponding pure solutal steady states,
since $E_C$ is negative) and the upper branch $E_+$ is thermal.
For $-1/L < S < -L$, the pitchfork bifurcation occurs at negative $\rpf$
and branches towards lower $r$; the resulting branches are solutal.
(These do correspond to pure solutal steady states.)
The branches that exist for $r>\rsn$ are isolated. The high-amplitude 
branches $E_+$ are primarily thermal and the low-amplitude branches 
$E_-$ solutal.
For $S<-1/L$, the isolated branches no longer exist, leaving
only the solutal branches, whose amplitude increases with $\vert S \vert$.

We now focus on positive $S$. 
Although less studied than negative $S$, 
this case has nonetheless received substantial attention for the Soret problem.
Qualitatively, in experiments or three-dimensional calculations, a striking 
pattern of squares is produced
\cite{Legal,Moses86,Silber,MullerLucke88,Knobloch89,Moses91,CluneKnobloch,Dominguez,LuckeBig,Jung,Huke},
possibly alternating with rolls of different orientation
Quantitatively, a fairly abrupt transition is observed 
between a low-amplitude and a high-amplitude convective regime,
as can be seen in figures \ref{fig:squareroot}.
This abrupt transition was first derived for the five-variable model
by Platten and Chavepeyer \cite{PlattenChavepeyer76}, 
first observed experimentally by Le Gal et al. \cite{Legal}, 
and the two regimes identified and named the Soret and Rayleigh regimes 
by Moses and Steinberg \cite{Moses86,Moses91}.
The Soret-to-Rayleigh transition has also been reproduced experimentally 
in \cite{Ahlers,LhostPlatten,Dominguez}
and numerically in \cite{LhostPlatten,KnoblochMoore90b,BartenBig,Bergeon}.
We will interpret this transition 
as a manifestation of the avoided crossing phenomenon which occurs
at the separation between the solutal and thermal regimes.

The physically significant real and positive values of $E$
are those for $r>\rpf$. 
Thus, a steady state branch has physically significant
solutal and thermal portions 
if $\rintnl > \rpf$, which occurs if $L^3 < S < 1/L$.
In this case, the solutal regime comprises:
\begin{equation}
\frac{L}{L+S}  = \rpf < r < \rintnl = \frac{1-L^2}{1-LS}
\end{equation}
The transition between solutal and thermal steady states
is manifested by the related increase in slope (for $E$)
or curvature (for $A$).
For $S \geq 1/L$, the entire solution branch is solutal 
while for $0 < S \leq L^3$ the entire solution branch is thermal 
(see figure \ref{fig:nonplus}).

\begin{figure}
\vspace*{-2cm}
\centerline{
\psfig{file=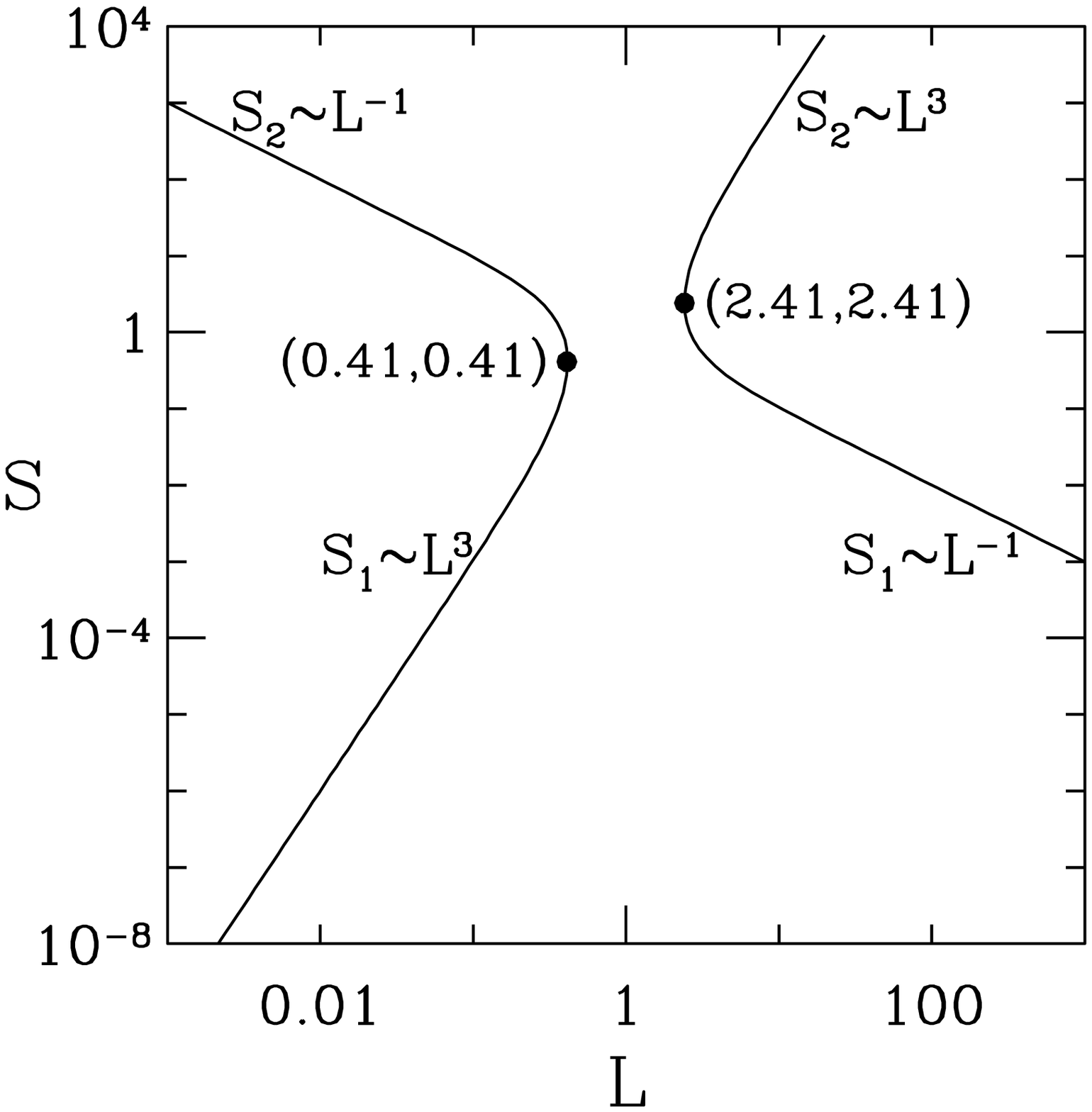,width=8cm}}
\caption{{}
Range $(S_1,S_2)$ of values for which $\rmidnl > \rpf$,
as a function of $L$.
This range exists for $L < 0.41$ or if $L>2.41$.}
\label{fig:soretrange}
\vspace*{-0cm}
\centerline{
\psfig{file=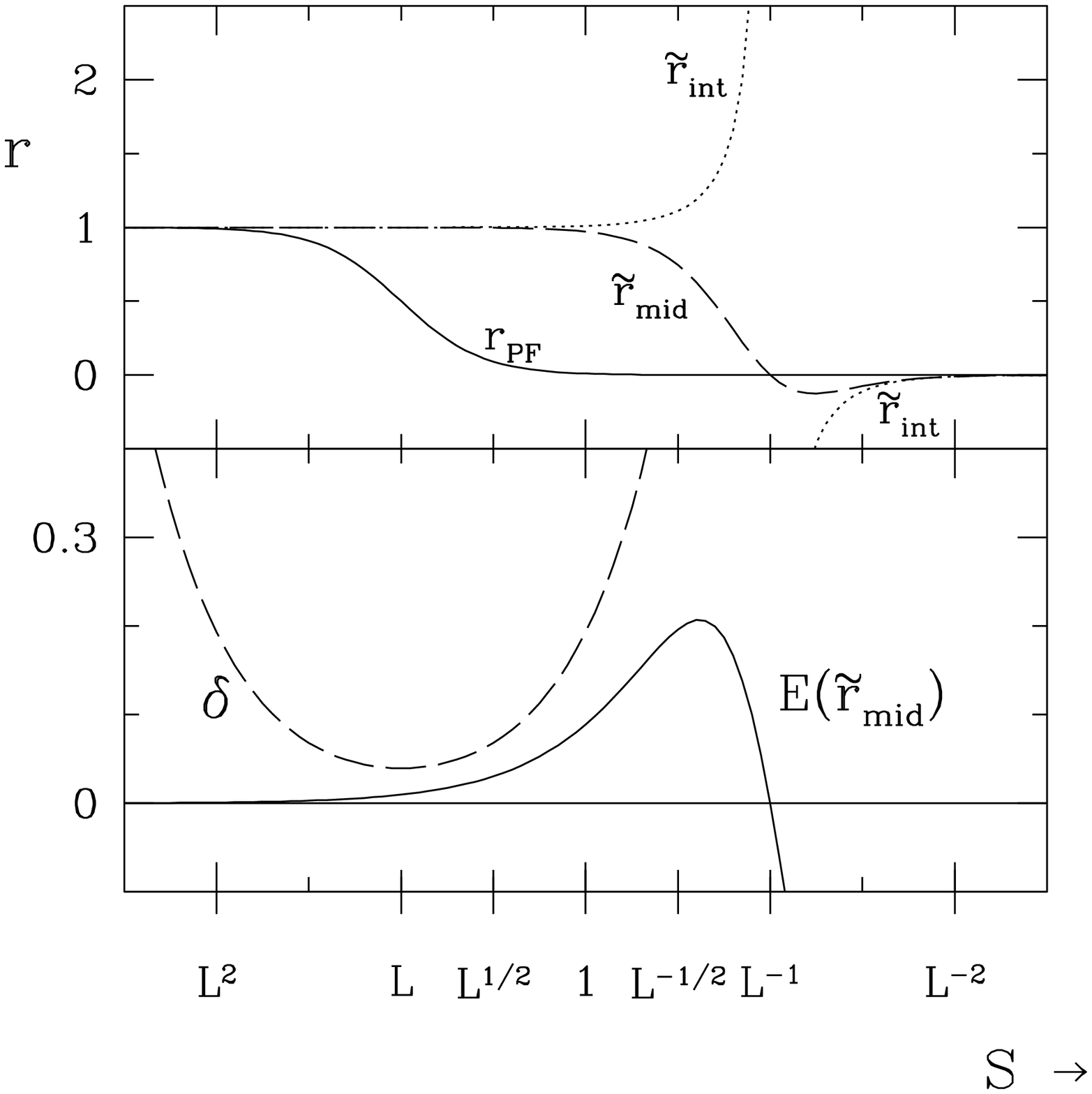,width=12cm}}
\vspace*{-0cm}
\caption{{}
Quantities determining existence of Soret-to-Raleigh-regime
transition for positive $S$.
Upper diagram:
Soret regime exists for $\rpf < r < \rmidnl$.
$\rmidnl-\rpf$ is positive for $L^3 < S < 1/L$ and
appreciable for $L^{5/4} \lesssim S \lesssim L^{-3/4}$.
Lower diagram:
Abruptness of transition is measured by the smallness of the ratio
$\delta$ (dashed curve) defined in (\ref{abrupt}) 
which remains less than 0.2 for
for $L^2 \lesssim S \lesssim 1$.
Energy $E(\rmidnl)$ at transition (solid curve) is maximal at $1/(4L)$ and
non-negligible for $L \lesssim S \lesssim 1/L$.
}
\label{fig:soretstats}
\end{figure}

What is actually observed is 
more complicated than the analysis given above.
Recall that the asymptotes of the hyperbolas (\ref{hyperbolanon}) 
describing $E$ differ from the lines $E_T(r)$ and $E_C(r)$.
The increase in slope of $E$ occurs, not at the intersection
point $\rintnl$ between the pure solutal and thermal branches,
but at the intersection point $\rmidnl$ between the two asymptotes; 
see Appendix \ref{Conic sections and eigenvalues}.
For $0<S<1$ and for $S > 10^4$, $\rmidnl$ and $\rintnl$ are so 
close as to be indistinguishable 
on figure \ref{fig:soretstats}.
The same proviso holds for $\rmidnl$ as for $\rintnl$: 
the transition 
is observed only if it occurs at a real and positive value of $E$, 
i.e. only if $\rmidnl > \rpf$.
In this case, the Soret regime comprises:
\begin{equation}
\frac{L}{L+S}  = \rpf < r < \rmidnl = \frac{(1-L^2)(1-LS)}{(1+LS)^2}
\label{rpf-rmidnl}
\end{equation}

For $S=0$, we have $\rmidnl = 1-L^2 < 1 = \rpf$, so the
entire solution branch is in the Rayleigh regime.
For $S$ large, it can also be shown that $\rmidnl < \rpf$;
the entire branch is then in the Soret regime.
But there is an intermediate range of $S$ over which the transition
can be observed, as can be seen from 
setting $\rpf$ equal to $\rmidnl$:
\begin{eqnarray}
\frac{L}{L+S}  &=& \frac{(1-L^2)(1-LS)}{(1+LS)^2}\nonumber\\
0 &=& LS^2 - (1 - 4L^2 + L^4)S + L^3 \nonumber\\
S_{1,2} &=& 
\frac{1}{2L} \left[ 1 - 4L^2 + L^4 \pm (1-L^2)\sqrt{1-6L^2 +L^4} \right]
\label{soretrange}
\end{eqnarray}
Equation (\ref{soretrange}) has two real solutions $S_1, S_2$ if 
$1-6L^2 +L^4$ is positive,
which occurs if $L < \sqrt{3-2\sqrt{2}} = 0.41$ or if 
$L > \sqrt{3+2\sqrt{2}} = 2.41$, as illustrated in figure 
\ref{fig:soretrange}.
For $L$ satisfying these conditions, and for $S_1 < S < S_2$,
there exists a low-amplitude Soret regime over interval (\ref{rpf-rmidnl}).
For $L \ll 0.41$, we have $S_1 \sim L^3$ and $S_2 \sim 1/L$
and vice versa for $L \gg 2.41$.
(Recall that $\rintnl > \rpf$ for $L^3 < S < 1/L$.)
Figure \ref{fig:soretstats} shows the variation with $S$ of
$\rmidnl$, $\rpf$, and $\rintnl$ for the case $L=0.01$.
We see that while the size of the Soret range $\rmidnl-\rpf$ is 
positive for $L^3 \lesssim S \lesssim 1/L$, it is appreciable only 
over the smaller interval $L^{5/4} \lesssim S \lesssim L^{-3/4}$.

For the transition to be observed, the energy must also
be sufficiently large for convection to be detected.
\begin{equation}
E(\rmidnl) = \Emid+\sqrt{\Sgen} = \frac{-L(S+L)+\sqrt{SL}(1-L^2)}{1+LS}
\label{Ermidnl}
\end{equation}
is maximal at $S=(1-L)^2/(4L) \approx 1/(4L)$ for $L\ll 1$
and is negative or small outside the range $L lesssim S \lesssim 1/L$
(see, e.g., figure \ref{fig:nonplus} for S=0.001).

Another factor that blurs the transition from 
Soret to Rayleigh regime is the fact that
as $S$ increases above 0, the hyperbola (\ref{hyperbolanon}) 
separates from its 
asymptotes, and the change in slope becomes more gradual as seen,
for example, in figure \ref{fig:nonplus} for S=10.
(This is somewhat counterbalanced by the fact that the angle
between the two asymptotes increases from $\pi/4$
at $S=0$ to $\pi/2$ at $S=\infty$.)
The change in slope of $E$ is measured by $E''$,
which is maximal at $\rmidnl$ 
(see Appendix \ref{Conic sections and eigenvalues}).
Normalizing $E''(\rmidnl)$ by $E(\rmidnl)$ and
taking the inverse square root defines a length in $r$ 
over which the change in slope occurs.
Dividing this length by $\rmidnl-\rpf$ yields a ratio which compares this
length to the extent of the Soret regime.
Thus we define
\begin{equation}
\delta \equiv \frac{1}{\rmidnl-\rpf}
\left[\frac{E(\rmidnl)}{E''(\rmidnl)}\right]^{1/2} 
\label{abrupt}\end{equation}
where
\begin{equation}
E''(\rmidnl) = \frac{(1+LS)^3}{4\sqrt{SL}(1-L^2)} 
\end{equation}
and $\rpf$, $\rmidnl$, and $E(\rmidnl)$ are given in (\ref{rpf-rmidnl}) and 
(\ref{Ermidnl}). 
Figure \ref{fig:soretstats} shows that $\delta$ is
smallest -- i.e., the change is most abrupt -- near $S=L$ and 
rises steeply for $S\gtrsim 1$ and $S \lesssim L^2$.
The interplay of counterbalancing criteria demonstrates the 
multiple roles played by $S$ in this geometric analysis.

Combining all of these criteria,
we finally obtain $L \lesssim S \lesssim 1$
as the separation parameter range for
the Soret-to-Rayleigh transition to be observable.
The experimental observations of the Soret regime 
have indeed been approximately in this range.
Moses and Steinberg \cite{Moses91},
who have carried out the most extensive experimental investigation,
observe the Soret regime
for $L^{0.85} < S < L^{0.27}$. Other experimental observations
are at $S=L^{1.12}$ \cite{Ahlers}, at $S=L^{1.15}$ \cite{LhostPlatten},
and at $S=L^{0.99}$ \cite{Dominguez}.
Numerical observations should be possible over a larger range:
since $r$ and $E$ ranges can effectively be magnified as required,
the thresholds for $\rmidnl -\rpf$ and $E(\rmidnl)$ are not as
constraining. Platten and Chavepeyer \cite{PlattenChavepeyer76}
observe the Soret regime for $L^{1.96} < S < L^{0.5}$,
other subsequent investigators reported a transition 
at $S=L^{1.13}$ \cite{Bergeon},
at $S=L^{0.21}$ \cite{LhostPlatten},
at $S=L^{1.46}$ \cite{KnoblochMoore90b},
and at $S=L^{0.5}$ \cite{BartenBig}.
For all values $S<1$, the transition point $\rmidnl$ 
is indistinguishable from the
thermal threshold $r_T = 1$ (see figure \ref{fig:soretstats}) and thus the
Soret-to-Rayleigh transition is invariably described as coinciding
with the onset of thermal convection in a pure fluid.

All of the references cited above have investigated the Soret problem.
There, the no-flux boundary conditions on $C$ lead to $k=0$ as
yielding the lowest threshold for linear instability
for sufficiently large $S$, e.g. \cite{Nield,KnoblochMoore88}.
This zero-wavenumber instability 
is sometimes invoked as part of the explanation 
for the weak heat transport in the
Soret regime \cite{LhostPlatten,Moses91}.
However, the wavenumber actually realized in 
full nonlinear simulations \cite{BartenBig} 
for the Soret problem with rigid boundaries 
in a two-dimensional geometry is close to $\pi$,
as is the wavenumber for the square patterns observed 
experimentally, e.g. \cite{Moses91,Dominguez}.
Considerations of pattern selection clearly play no role in 
the mechanism we have discussed since, in the thermosolutal problem,
the boundary conditions on $T$ and $C$ are identical and we
have fixed $k$ at $k_{\rm crit}=\pi/\sqrt{2}$.

Finally, we consider the asymptotic behavior of the thermosolutal 
solution branches.
For $|r|$ large, we have, for the upper branch $E_+$,
\begin{subequations}\begin{eqnarray}
\frac{SC}{T} &\approx& LS {\rm ~~with~~} |LS| 
\lessgtr
\sqrt{|LS|} {~~\rm if~~} |S| 
\lessgtr
1/L \label{TSCupper} \\
\frac{E}{E_T} &\approx& \frac{(1+LS)r}{r-1} \approx 1+LS \approx 
\left\{ \begin{array}{c} 1 \\ LS \end{array}\right\} 
{\rm ~~if~~} \vert S \vert 
\left\{\begin{array}{c} \ll \\ \gg \end{array}\right\} 
1/L \label{eet+}\\
\frac{E}{E_C} &\approx& \frac{(1+LS)r}{LS(r-S/L)} \approx \frac{(1+LS)}{LS}
\approx \left\{ \begin{array}{c}
1/(LS) \\ 1 \end{array}\right\} {\rm ~~if~~} \vert S \vert 
\left\{\begin{array}{c} \ll \\ \gg \end{array}\right\} 1/L \label{eec+}
\end{eqnarray}\end{subequations}
Relation (\ref{TSCupper}) shows that the upper branch
is thermal if $|S| < 1/L$ and solutal if $|S| > 1/L$.
Relations (\ref{eet+}-\ref{eec+}) suggest another, 
more stringent, criterion for classification: 
a steady state is thermal (solutal) if the
ratio $E/E_T$ ($E/E_C$) is close to one, 
which is true for the upper branch if $|S| \ll (\gg) 1/L$.

For the lower branch $E_-$,
\begin{subequations}\begin{equation}
\frac{SC}{T} \approx -1 
{\rm ~~with~~} 1
\gtrless
\sqrt{|LS|} {~~\rm if~~} |S| 
\lessgtr
1/L
\label{TSClower}
\end{equation}
so the lower branch 
is solutal if $|S| < 1/L$ and thermal if $|S|> 1/L$,
by the criterion (\ref{proximitynon})
but it fails to meet the more stringent criterion
since the ratios $E/E_T$ and $E/E_C$ both tend to zero:
\begin{eqnarray}
\frac{E}{E_T} &\approx& \frac{-L(L+S)/(1+LS)}{r-1} \\
\frac{E}{E_C} &\approx& \frac{-L(L+S)/(1+LS)}{LS(r-S/L)} 
\end{eqnarray}\end{subequations}

\begin{figure}
\vspace*{-1cm}
\centerline{
\psfig{file=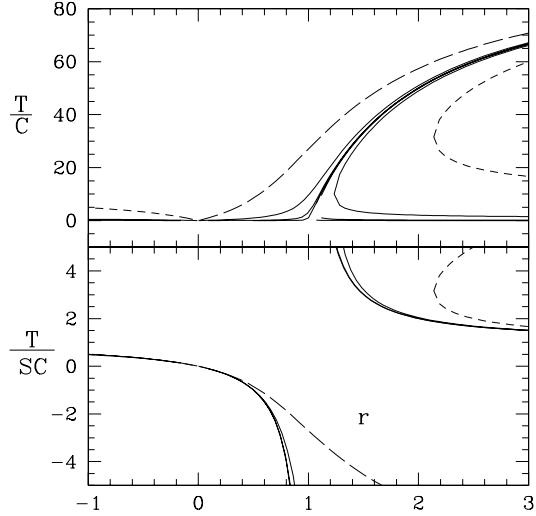,width=8cm}}
\caption{{}
Solid curves show $T/C$ (above) and $T/(SC)$ (below) for 
$S=\pm 1, \pm 0.1, \pm 0.01$;
long- (short-) dashed curves show $T/C$ or $T/(SC)$ for $S= +(-)10$.
As $r$ becomes large (and along the upper of the two branches for $S$ negative), 
$T/C$ approaches its asymptotic value of $1/L$, independent of $S$.
For low amplitude states existing for $S$ negative, $T/(SC)$ approaches $-1$.
}
\label{fig:ratios}
\end{figure}

The asymptotic behavior (\ref{TSCupper}) and (\ref{TSClower})
is illustrated in figure \ref{fig:ratios}.

\section{Time-dependent model}
The linear and nonlinear equations 
of sections \ref{Linear analysis} and \ref{Nonlinear analysis} :
\begin{subequations}\begin{eqnarray}
\frac{d}{dt}\left(\begin{array}{c} T \\ C \end{array}\right)
&=&\left(\begin{array}{cc} r-1 & Sr \\ r & Sr-L \end{array}\right)
\left(\begin{array}{c} T \\ C \end{array}\right) \label{hybridlin}\\
\left(\begin{array}{c} 0 \\ 0 \end{array}\right)
&=&\left(\begin{array}{cc} r-1 & Sr \\ r & Sr-L \end{array}\right)
\left(\begin{array}{c} T \\ C \end{array}\right)
-\frac{1}{2}\left(\frac{r\geom}{2}\right)^2 (T+SC)^2
\left(\begin{array}{cc} 1 & 0 \\ 0 & 1/L \end{array}\right)
\left(\begin{array}{c} T \\ C \end{array}\right) \label{hybridnon}
\end{eqnarray}\end{subequations}
each lack an essential feature
of the bifurcation diagram for binary fluid convection:
the linear stability problem (\ref{hybridlin}) cannot contain 
saddle-node bifurcations, whereas the nonlinear steady-state 
problem (\ref{hybridnon}) cannot describe Hopf bifurcations.
We can combine (\ref{hybridlin}) and (\ref{hybridnon}) to form a single 
time-dependent two-variable system containing all of these features:
\begin{equation}
\frac{d}{dt}\left(\begin{array}{c} T \\ C \end{array}\right)
=\left(\begin{array}{cc} r-1 & Sr \\ r & Sr-L \end{array}\right)
\left(\begin{array}{c} T \\ C \end{array}\right)
-\frac{1}{2}\left(\frac{r\geom}{2}\right)^2 (T+SC)^2
\left(\begin{array}{cc} 1 & 0 \\ 0 & 1/L \end{array}\right)
\left(\begin{array}{c} T \\ C \end{array}\right)
\label{hybrid}\end{equation}
The conditions for the validity of this two-dimensional time-dependent 
nonlinear system combine those required for the two
systems (\ref{hybridlin}) and (\ref{hybridnon}): 
large Prandtl number and small amplitudes.
In addition, the spatial representation (\ref{solform})
imposes a fixed phase on the solutions.
Beyond these two statements, we make no claim for the accuracy of
system (\ref{hybrid}) as a representation of the partial
differential equation (\ref{thermosolutal_full}).

\begin{figure}
\vspace*{-2cm}
\centerline{
\psfig{file=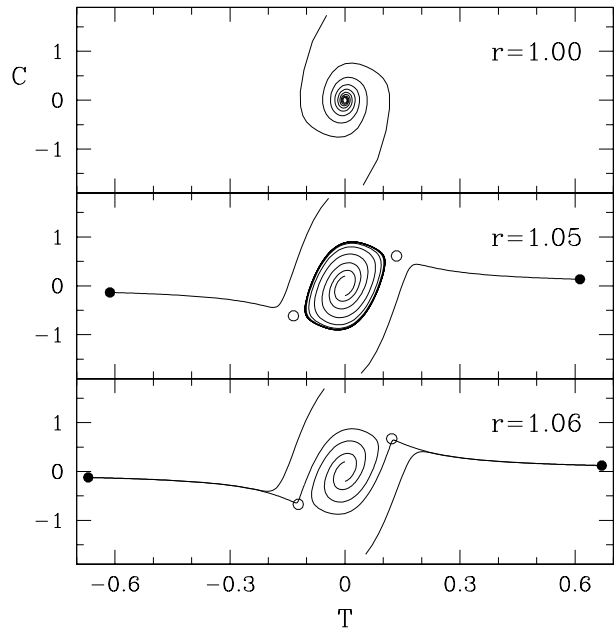,width=9cm}}
\caption{{}
Phase portraits illustrating heteroclinic bifurcation for $S=-L=-0.01$.
For $r=1.00$, all trajectories spiral into $(T,C)=(0,0)$.
For this value of $S$, the Hopf and saddle-node bifurcations
occur simultaneously at $r=1.02$.
For $r=1.05$, trajectories originating close to zero spiral out
to a limit cycle, while trajectories originating sufficiently far from
zero terminate on one of the stable steady states (solid dots), possibly
after being deflected by one of the saddle points (hollow dots).
For $r=1.06$, the limit cycle has been destroyed by colliding
with the saddle points in a heteroclinic bifurcation and all trajectories
terminate on one of the stable steady states.
}
\label{fig:phase}
\end{figure}

By construction, system (\ref{hybrid}) 
undergoes a pitchfork bifurcation at $r = \rpf = L/(L+S)$ and 
a Hopf bifurcation at $\rh = (1+L)/(1+S)$, both from the
trivial state, and 
saddle-node bifurcations at $r =\rsn = (1-L^2)/(1-\sqrt{-LS})^2$.
In addition to reproducing the linear stability of the conductive
state and the nonlinear steady states, this system also displays
an interesting phenomenon that can occur in actual binary fluid convection:
the limit cycle disappears via a global bifurcation by
colliding with the saddles on the lower branch of steady states.
In figure \ref{fig:phase}, we show numerically computed
phase portraits of (\ref{hybrid}) for $S=-L=-0.01$.
Initial conditions are $(T,C)=\pm(0,0.2)$, $\pm (0,2.0)$.
For this value of $S$, the Hopf and saddle-node bifurcations
occur simultaneously at $r=1.02$.
For $r$ slightly less than $r=1.02$, the origin is
a stable spiral node; all trajectories spiral into (0,0).
At $r=1.02$, a limit cycle and two pairs of steady states
(stable nodes and unstable saddles) are created.
For $r$ slightly exceeding $r=1.02$, trajectories
approach either the limit cycle or the stable steady state,
depending on the initial condition.
At $r=r_G$ with $1.05 < r_G < 1.06$, the heteroclinic bifurcation
destroys the limit cycle.
For $r \gtrsim 1.06$, all trajectories terminate on one
of the stable steady states.
The influence of the saddles on the trajectories is clearly seen.

\begin{figure}
\vspace*{-5cm}
\centerline{
\psfig{file=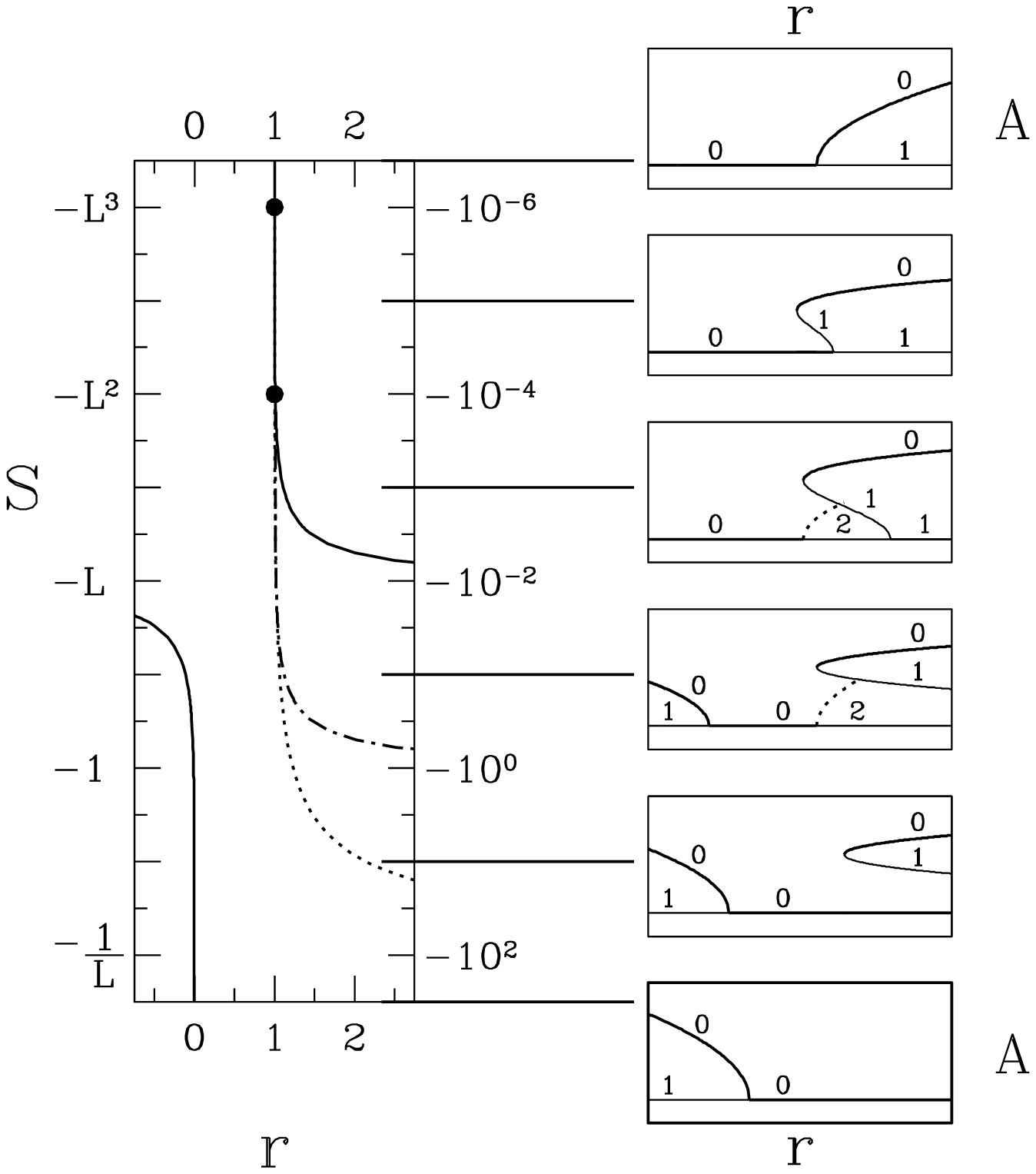,width=20cm}}
\vspace*{-1cm}
\caption{{}
Behavior of the hybrid model for negative $S$. \newline
Left: Thresholds for negative values of $S$ plotted on a logarithmic scale.
Solid curves show the thresholds $\rpf=L/(L+S)$ of pitchfork bifurcations, 
negative for $S<-L$ and positive for $S>-L$.
Dash-dotted curve indicates the thresholds $\rh = (1+L)/(1+S)$ of 
Hopf bifurcations, 
appearing from $r=\infty$ at $S=-1$ and disappearing by meeting the
pitchfork bifurcation curve in a codimension-two (Bogdanov) bifurcation
at $S = -L^2$ (large dot).
Dotted curve indicates the thresholds $\rsn$ of saddle-node bifurcations,
appearing from $r=\infty$ at $S=-1/L$ and disappearing by meeting
the pitchfork bifurcation curve in a codimension-two (degenerate pitchfork) 
bifurcation at $S=-L^3$ (large dot).\newline
Right: Schematic bifurcation diagrams for the six qualitatively
different cases. 
Numbers above branches indicate the number of eigenvectors to
which the branch is unstable. Stable branches (0 unstable
eigenvectors) are additionally shown as heavy curves.
Dotted curves indicate limit cycles.
For $S<-1/L$, the pitchfork bifurcation is supercritical and 
branches backwards from negative $\rpf$.
For $-1/L < S < -1$, additional disconnected branches are
created via saddle-node bifurcations.
For $-1 < S < -L$, a Hopf bifurcation creates a limit
cycle which terminates via a global heteroclinic bifurcation.
For $-L < S < -L^2$, the supercritical pitchfork bifurcation 
branches backwards from positive $\rpf$ and connects the branches
emanating from the saddle-node bifurcations.
For $-L^2 < S < -L^3$, the Hopf bifurcation no longer exists
and the pitchfork bifurcation is now subcritical.
For $-L^3 < S$, the saddle-node bifurcations no longer exist
and the pitchfork bifurcation is again supercritical.\newline
}
\label{fig:hybrid}
\end{figure}

In figure \ref{fig:hybrid} we show the thresholds for the
bifurcations undergone by model (\ref{hybrid}). 
The accompanying bifurcation diagrams are schematic, unlike
those of figures \ref{fig:linminus}, \ref{fig:linplus}, 
\ref{fig:nonminus}, and \ref{fig:nonplus}.
There are six qualitatively different
diagrams, for $S$ in ranges $S < -1/L$, $-1/L < S < -1$, 
$-1 < S < -L$, $-L < S < -L^2$, $-L^2 < S < -L^3$, and $-L^3 < S$.
This illustrates the advantage of the simplified model:
each change in the qualitative dynamics occurs exactly at a power of $L$.
Each steady branch is labeled with the number of eigenvectors to 
which it is unstable, i.e. the number of eigenvalues with 
positive real part.
In section \ref{Nonlinear analysis}, we characterized the
pitchfork bifurcations as forward or backward facing,
according to whether the new solutions created
branch towards $r>\rpf$ or $r<\rpf$,
rather than as supercritical or subcritical.
The reason for this is that the criterion we use for super or 
subcriticality depends on a combination of linear and nonlinear
information: 
a bifurcation is supercritical if the new solutions
branch in the direction of increasing instability of
the parent branch.
In this sense, the pitchfork bifurcations 
are supercritical for all cases except $-L^2 < S < -L^3$, 
despite being backward facing for the five cases $S < -L^3$.
For the three cases $S < -1/L$, $-1/L < S < -1$, and
$-1 < S < -L$, a real eigenvalue becomes and
remains positive as $r$ is decreased below $\rpf$.
For the fourth case $-L<S<-L^2$, the positive eigenvalue coalesces
with another positive eigenvalue to form a complex conjugate
pair whose real part then reverses direction and becomes negative 
as $r$ is decreased, resulting in the Hopf bifurcation at $\rh$,
as was shown in figure \ref{fig:linminus} for $S=-0.001$.
Although we have shown the Hopf bifurcation as supercritical
in figure \ref{fig:hybrid}, we do not exclude the possibility
of a subcritical Hopf bifurcation accompanied by a stabilizing 
saddle-node bifurcation.

In a two-variable system, the heteroclinic bifurcation 
is a natural consequence of Hopf bifurcations and
pitchfork bifurcations which branch towards each other.
Starting from the Hopf bifurcation and 
approaching the pitchfork bifurcation, the amplitude of 
the limit cycle increases while that of the steady branch 
(of saddle points) decreases.
Confined to a plane, the limit cycle and steady states then 
collide at some intermediate value of $r$.

More generally, this bifurcation sequence has been studied extensively  
\cite{HuppertMoore,KnoblochProctor,DaCosta,GuckenheimerKnobloch,KnoblochMooreToomreWeiss,Rucklidge,KnoblochProctorWeiss}.
The approach to the heteroclinic bifurcation has been observed in
numerical simulations of the full system of governing partial differential 
equations for thermosolutal convection \cite{HuppertMoore,KnoblochMooreToomreWeiss},
and for the Marangoni-Soret problem \cite{Bergeon}.
The existence of the heteroclinic bifurcation has been proven 
using the techniques of normal form reduction and amplitude expansions, 
and analytic expressions calculated for the limit cycles and the bifurcation
\cite{KnoblochProctor,GuckenheimerKnobloch,KnoblochProctorWeiss}.
In addition, simulations of the five-mode Veronis model exhibit interesting 
complex dynamical phenomena such as period-doubling and chaos \cite{DaCosta}; 
this is not possible for a two-variable model such as (\ref{hybrid}).
Mathematical analyses have rigorously derived reduced models of binary fluid 
convection, analyzed their domains of validity, and determined 
when period-doubling and chaos may occur \cite{Rucklidge,KnoblochProctorWeiss}.

Referring to (\ref{solform}), the limit cycle of
figure \ref{fig:phase} is a standing wave solution of the 
partial differential equations (\ref{thermosolutal_full}).
In treatments of (\ref{solform}) adapted to large or
infinite horizontal domains, in which
the phases of the various components are allowed to vary,
then the Hopf bifurcation gives rise to
a branch of stable traveling waves which disappears via a 
drift bifurcation by meeting the branch of stable steady states;
the standing waves described above continue to exist but are unstable 
\cite{BrethertonSpiegel,Knobloch85,CoulletFauveTirapegui,Knobloch86}.
Traveling waves are indeed observed in containers which are large
or periodic in the horizontal direction \cite{Walden,Deane1,Deane2,BartenBig}.
The situation is in fact far more complicated: 
the traveling or standing wave branches may bifurcate subcritically, e.g.
\cite{HuppertMoore,PlattenLegros,Knobloch86,Schopf},
the traveling wave branch undergoes a secondary bifurcation
to modulated traveling waves \cite{Knobloch86,Deane2,KnoblochMoore90}, 
and, at least for the Soret problem, the traveling wave branch
can undergo several saddle-node bifurcations between
slow and fast branches \cite{HollingerBuchelLucke}.
As a further complication, in larger two-dimensional domains, 
localized traveling waves and pulses 
predominate \cite{Moses87,Swinney,BartenBig2}.
Yet, certain large-scale aspects of steady-state 
convection examined in section \ref{Nonlinear analysis}
have counterparts for standing and traveling waves.
Sch\"opf and Zimmerman \cite{Schopf} have found
that the degenerate Hopf bifurcations for the standing and 
traveling wave branches are located at $S \sim -L^2$.
Hollinger et al. 
\cite{HollingerBuchelLucke,HollingerLucke98,HollingerLuckeMuller}
have shown that $r$ is a simple function
of the amplitude and frequency of traveling waves 
and that the traveling wave branch can be divided into low-amplitude 
Soret and high-amplitude Rayleigh regimes.

Many attempts have been made to reduce the governing-fluid 
dynamical equations to minimal models which describe traveling waves.
The most obvious approach is to extend the five-mode free-slip Veronis 
model (\ref{full5Dmodel}) to include additional modes proportional to 
$\sin(kz)$ \cite{Cross86,AhlersLucke}.
However, the resulting eight-mode model proves to be singular, because
the Hopf bifurcation to traveling waves in binary fluid convection 
with free-slip boundary conditions is always degenerate 
\cite{BrethertonSpiegel,Knobloch85,Knobloch86,Deane1,Swinney},
as a consequence of (\ref{preserve}).
This means that the truncation (\ref{solformphi}), (\ref{solform_full})
is insufficient for even a qualitative description of traveling
waves in the full thermosolutal problem (\ref{thermosolutal_scaled}).
Numerous other models have ensued 
\cite{LinzLucke87,Bensimon,KnoblochMoore90b,KnoblochMoore90,Riecke,Schopf,HollingerLuckeMuller},
using other boundary conditions, additional field variables
or different theoretical approaches.
Understanding the diverse aspects of traveling waves 
in binary fluid convection is an extremely challenging problem.
\section{Conclusions}
We have examined the well-known idealized thermosolutal problem from 
a variety of different perspectives.

For infinite or large Prandtl number, the linear stability problem 
for temperature and concentration perturbations $(T, C)$ is governed 
by a $2 \times 2$ matrix
$M$ whose entries depend linearly on the reduced Rayleigh number $r$,
and whose eigenvalues provide the growth rates of
perturbations to the motionless conductive state.

We interpret the diagonal terms of this thermosolutal matrix as
growth rates of two ``pure'' convection problems driven
exclusively by a thermal gradient or by a concentration gradient,
which we term the thermal and the solutal eigenvalues.
Without coupling, the thermosolutal eigenvalues merely cross
transversely as $r$ is varied through the intersection point $\rint$ 
of the thermal and the solutal eigenvalues.
Otherwise, the two eigenvalues of the coupled problem either undergo 
avoided crossing (the eigenvalues appear to deflect each other
and remain real) or complex coalescence (the two eigenvalues
join into a complex conjugate pair and then become real again).
Which possibility is realized depends on the
sign of the coupling: the product of off-diagonal terms.
For the thermosolutal problem the coupling is proportional
to, and has the same sign as, the separation parameter $S$.

In the equivalent language of conic sections, 
the eigenvalues $\sigma + i \omega$ and reduced Rayleigh number $r$
satisfy a second-degree equation.
Hence the sets $(r,\sigma)$ and $(r,\omega)$ form
hyperbolas, parabolas, or ellipses, according to the values
of two invariants. The first is the discriminant
of the matrix responsible for advection,
which here is positive (except for the single value $S=-1$).
The curves $(r,\sigma)$ are thus hyperbolas and the
curves $(r,\omega)$ ellipses.
The second invariant $\Delta$ is, 
for the thermosolutal problem, 
proportional to the separation constant $S$.
$\Delta=0$ is the limiting case of a hyperbola
consisting of two intersecting lines, or of
an ellipse whose radii are zero.
As $\Delta$ changes sign, the quadrants occupied
by the hyperbola shift, and the ellipse becomes empty.

Both of these equivalent formulations underline the 
organizing role played by the delimiting case $S=0$.
Most studies of binary fluid convection treat the codimension-two
point \cite{KnoblochProctor} 
$S_* = -L^2$, where the pitchfork and Hopf bifurcation 
curves meet at $\rpf = \rh = r_* = (1-L)^{-1}$,
as a distinguished point in the $(S,r)$ plane, and expand around it. 
Our complementary point of view focuses on $S=0$, where
the pure thermal and solutal eigenvalues intersect at $\rint = 1-L$,
as a different kind of distinguished point.

Turning to the nonlinear problem, 
the minimal model of thermosolutal convection incorporating 
the lowest-order nonlinear effects was first derived by 
Veronis \cite{Veronis65} and has since been extensively studied.
We find that the system of nonlinear equations satisfied by the 
steady states of the minimal model is of the special form:
\begin{equation}
\tilde{M} v = E(v_1, v_2, ... ) v
\label{special}\end{equation}
where $\tilde{M}$ is a matrix, $v$ a vector, and $E$ a scalar function 
of the components of $v$.
The solution of such systems reduces to that of
diagonalizing a matrix and solving a single nonlinear
equation of one variable.
The eigenvalues of $\tilde{M}$ are the possible values of $E$.
Its eigenvectors are used to reduce the number of arguments of $E$
to one by rewriting all but one component of $v$
as multiples of the remaining component.

The five-variable system governing the steady states of the
minimal thermosolutal model can be further reduced to a two-variable 
system in $(T,C)$ of type (\ref{special}) in which the scalar function 
$E$ is proportional to the kinetic energy 
and the $2\times 2$ matrix $\tilde{M}$ bears a 
striking resemblance to the linear stability matrix $M$.
This leads to a remarkable analogy between the
linear stability problem and the nonlinear steady state problem.
The energy $E$ also undergoes avoided crossing or complex
coalescence, again according to the sign of $S$.
The curves $(r,E)$ are hyperbolas.
Complex coalescence for $E$ must be interpreted as the
disappearance of solution branches, rather than as the
onset of oscillatory behavior.
Quantitative results concerning the growth rates of perturbations
to the conductive state can be translated to results concerning 
the kinetic energy of nonlinear steady states
merely by transforming $S \rightarrow LS$ and $L \rightarrow L^2$,
where $L$ is the Lewis number, the ratio of solutal to
thermal diffusivities.

For some results -- the pitchfork bifurcation $\rpf$ -- 
the analogy leaves the linear result unchanged.
This reflects the bifurcation-theoretic fact that 
a change in sign in eigenvalue signals a bifurcation,
i.e. an intersection between solution branches.
For other results -- the Hopf bifurcation $\rh$ --
the analogy is valid, but not meaningful for the
nonlinear problem.
For yet other results -- the point at which eigenvalues
become complex $r_+$ and the codimension-two point $(S_*,r_*)$ --
the nonlinear analogues are both significant and different from 
the linear phenomena.
In particular, for negative $S$, the analogy indicates that there is a 
fundamental relationship between the onset of oscillation 
via the complex coalescence at $r_+$ of two real eigenvalues 
in the linear problem, and the termination of two 
finite-amplitude solution branches via saddle-node
bifurcation at $\tilde{r}_+ = \rsn $ for the nonlinear problem.
As a corollary, there also exists a fundamental relationship
between the merging of the pitchfork with the Hopf bifurcation
in the codimension-two point (Bogdanov bifurcation) at $(S_*,r_*)$
for the linear problem, 
and the merging of the pitchfork with the saddle-node bifurcation
in a different kind of codimension-two point (degenerate pitchfork)
at $(\tilde{S}_*,\tilde{r}_*)$ for the nonlinear problem.

The relationship is a consequence of the exact analogy between 
the growth rates of the linear stability problem 
at infinite $P$ and the energies of the nonlinear steady states 
of the minimal five-mode Veronis model both for
the thermosolutal problem with imposed horizontally periodicity 
and free-slip vertical boundaries.
We would not expect this analogy to persist  when any of these
assumptions are relaxed, i.e. when the boundary conditions are realistic,
when the concentration gradient occurs via the Soret effect, 
when the amplitudes or the Prandtl number are moderate rather than
infinitesimal or infinite.
Yet, evidence 
\cite{BainesGill,HurleJakeman,Brand,Knobloch86,Schopf,HollingerLuckeMuller}
suggests that the analogy must hold at least 
approximately, since the leading behavior of the two
codimension-two points, $S_* \sim -L^2$ and $\tilde{S}_* \sim -L^3$,
and the domains of existence of the Hopf and saddle-node bifurcations
continue to be related in approximately the same way 
as for the idealized thermosolutal problem.
This indicates that the analogy between growth rates and 
energies of steady states could be a fundamental unifying feature
of double-diffusive problems.

Another provocative feature of binary fluid convection is the fact that
the Rayleigh number is a single-valued function of any one of the following 
variables: growth rate, steady state energy,
traveling wave amplitude and frequency,
i.e. for any of these variables, each value is achieved at most once
when varying $r$.
We have explained this dependence for the growth rate and energy
by the way in which $r$ enters the problem via advection.
Hollinger et al. \cite{HollingerLucke98,HollingerLuckeMuller} 
provide a related explanation invoking 
the reduction of the velocity field to one mode and the resulting
simplification of the nonlinear terms.

Finally, we have proposed a classification of eigenvalues and
of steady states as primarily thermal or primarily solutal,
based on their proximity to the eigenvalues or steady states
of the pure thermal and pure solutal problems and on the 
relative proportions of solutal and thermal contributions 
to the buoyancy force.
For many $S$ values of interest, the coupling term is small
in the vicinity of the bifurcations, and so the convection 
threshold is very close to that of pure thermal or pure solutal convection.
The classification is particularly useful for the nonlinear problem 
for positive $S$.
We have shown that the relatively abrupt transition between small 
amplitude and large amplitude convection called the Soret and Rayleigh regimes
\cite{Legal,Ahlers,Moses86,LhostPlatten,Moses91,BartenBig,Dominguez}
corresponds to the change in 
slope seen as a hyperbola adheres to first one and then the 
other asymptote as $r$ is increased.
This transition may be masked, because it occurs
in a regime not corresponding to a real steady state,
or muted, because it is too gradual. We have been able to 
give precise conditions under which the transition from Soret to
Rayleigh regimes can be observed.

We emphasize that our goal has not been to reproduce 
all of the spatio-temporal dynamics of double-diffusive convection,
nor even to investigate the temporal dynamics of the five-mode
Veronis model.
Indeed, these goals have been admirably pursued and accomplished 
in previous research using other approaches, 
e.g. \cite{DaCosta,Knobloch86,KnoblochMoore90,Rucklidge,KnoblochProctorWeiss,Schopf,BartenBig,HollingerLucke98,HollingerLuckeMuller}.
Nor can this purpose be accomplished by the idealized
two-variable models we have investigated.
Instead, our goal has been to extract certain universal 
large-scale features of double-diffusive convection 
in as simple a context as possible, and to re-examine these
features in light of the avoided crossing/complex coalescence 
dichotemy and the linear/nonlinear analogy that we have put forth.

Convection in binary fluids has previously provided a testbed
for the discovery and realization of many fascinating phenomena 
in dynamical systems.
Our hope is that these new perspectives continue this tradition.

{\Large \bf Acknowledgments}

I am very grateful to  
Alain Bergeon and Daniel Henry for introducing me to the mysteries
of binary fluid convection.
I am indebted to Edgar Knobloch for his interest and encouragement and 
to Fritz Busse for suggesting the infinite Prandtl number limit.
I also thank Dwight Barkley, John Guckenheimer, Manfred L\"ucke, 
Ehouarn Millour, Hermann Riecke, and Alastair Rucklidge 
for helpful discussions and references.

\appendix
\parindent 0cm
\vspace*{1cm}
{\bf \LARGE Appendices}
\parindent .5cm
\section{Finite Prandtl number model}
\label{Finite Prandtl number model}

Although the results of section \ref{Nonlinear analysis} of 
concerning the nonlinear steady states are
independent of Prandtl number $P$, those of section 
\ref{Linear analysis} concerning the linear stability problem
are derived by taking $P$ to be infinite.
In this Appendix, we describe our reduction of the thermosolutal 
linear stability problem to a $2\times 2$ matrix in the
case when $P$ is finite.
The key step in our interpretation is to decompose the velocity
field into ``thermal'' and ``solutal'' velocity fields
induced by the thermal and concentration gradients, with
vertical components $\hat{w}_T$ and $\hat{w}_C$, respectively.
Referring to equations (\ref{thermosolutal_lin}),
the linearized equations governing this augmented set of fields are:
\begin{subequations}
\begin{eqnarray}
\dt \hat{T} &=& \hat{w}_T + \hat{w}_C + \geom^{-2}\lap \hat{T} \label{dec1}\\
\dt \geom^{-2}\lap \hat{w}_T &=& Pr k^{-2}\dx^2 \hat{T} + P\geom^{-4}\nabla^4 \hat{w}_T \label{dec2}\\
\dt \hat{C} &=& \hat{w}_T + \hat{w}_C + L \geom^{-2}\lap T \label{dec3}\\
\dt \geom^{-2}\lap \hat{w}_C &=& PSr k^{-2}\dx^2 \hat{C} + P\geom^{-4}\nabla^4 \hat{w}_C \label{dec4}
\end{eqnarray}
\label{dec5}
\end{subequations}
Using the spatial and temporal dependence and notation
defined in (\ref{solform})-(\ref{solformtime}),
(\ref{dec5}) becomes:
\begin{equation}
\mu \left( \begin{array} {c} T \\ w_T \\ C \\ w_C \end{array} \right) 
=
\left( \begin{array} {c c c c}
-1 & 1 & 0 & 1 \\
Pr & -P & 0 & 0 \\
0 & 1 & -L & 1 \\
0 & 0 & PSr & -P
\end{array} \right)
\left(\begin{array}{c}T\\w_T\\C\\w_C\end{array}\right) 
\label{decmatrix}
\end{equation}
This $4 \times 4$ system has exactly the same eigenvectors and eigenvalues
as system (\ref{thermosolutalmatrix}) with the additional eigenvector
$(T \; w_T \; C \; w_C) = 
(0 \; 1 \; 0 \; -1)$ and eigenvalue $\mu = -P$. 

Note that the temperature and concentration fields are each
advected by both the ``thermal velocity'' and the ``solutal velocity''.
It is this cross-advection which couples the thermal and solutal problem. 
Neglecting it leads to the decoupled thermal and solutal problems
discussed below.

\subsection{Thermal problem}

\begin{figure}[h]
\vspace*{0cm}
\centerline{\psfig{file=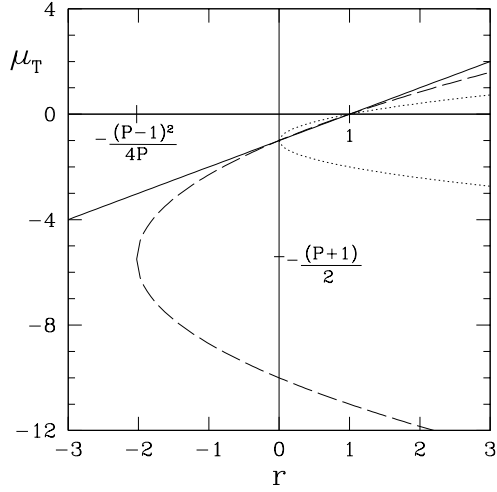,width=8cm}}
\vspace*{-.5cm}
\caption{{} Eigenvalues of pure thermal problem for $L=0.01$.
Dotted and dashed curves shows $\mu_T$ for $P=1$ and $P=10$,
respectively. Solid curve shows $\sigmat$, valid for $P=\infty$,
and also $\mu_T$ for $P=100$, from which it is indistinguishable
in this range. Coordinates of vertex are shown for $P=10$.}
\label{fig:thermeig}
\end{figure}

The upper left $2 \times 2$ submatrix of (\ref{decmatrix})
describes the onset of thermal convection in a simple fluid 
of finite Prandtl number.
This is perhaps the prototypical problem in hydrodynamic stability 
theory (e.g., \cite{Chandra,Gershuni,PlattenLegros}).
The corresponding thermal eigenvalues satisfy 
\begin{equation}
\mu_T \left( \begin{array} {c} T \\ w_T \end{array} \right) 
=
\left( \begin{array} {c c}
-1 & 1 \\
Pr & -P
\end{array} \right)
\left( \begin{array} {c} T \\ w_T \end{array}
\right) 
\label{purethermal}
\end{equation}
\begin{equation}
\mu_{T\pm} = \left[ -\left({{P+1}\over2} \right) \pm
\sqrt{\left({{P-1}\over2}\right)^2 + Pr} \right]
\label{purethermaleigs}\end{equation}
and are plotted in figure \ref{fig:thermeig}.

Figure \ref{fig:thermeig} shows that for $r < -(P-1)^2/(4P)$,
eigenvalues are complex; perturbations to the conductive
state oscillate as they decay.
(In the terms introduced in the appendix, the discriminant of
the $r$-dependent component of (\ref{purethermal}) is zero, 
so the curves of real values and
imaginary parts of $\mu_{T\pm}$ are both parabolas.)
We consider only the regime in which the eigenvalues are real.
Since we will only require the larger of the two eigenvalues,
we will write $\mu_T \equiv \mu_{T+}$.
For $P$ large, $\mu_T$ becomes $\sigmat$ of (\ref{sigT}),
as expected. 
The threshold of $\mu_T$ is $\rt= 1$, that of $\sigmat$.
Indeed, as is well known, the threshold of convection is
independent of $P$. The slope of $\mu_T$ at threshold is 
$P/(P+1)$, which also approaches that of $\sigmat$ for $P$ large.
We will also require the normalized right and left eigenvectors 
corresponding to $\mu_T$:
\begin{subequations}\begin{eqnarray}
{\mathcal T}^R \equiv \left( \begin{array}{c} T^R \\ w_T^R \end{array} \right)
= {1\over {N_T}} \left( \begin{array}{c} 1 \\ 
1 + \mu_T  \end{array} \right)
\label{righttherm} \\
{\mathcal T}^L \equiv \left( \begin{array} {cc} T^L & w_T^L \end{array} \right) = {{1}\over {N_T}} 
\left( \begin{array} {cc} P + \mu_T {\rm ~~}&{\rm ~~} 1 \end{array} \right)
\label{lefttherm}
\end{eqnarray}
where $N_T^2 \equiv 2\sqrt{((P-1)/2)^2 + Pr}$.
Thus
\begin{equation}
{\mathcal T}^L M_T {\mathcal T}^R = \mu_T
\label{diagtherm}
\end{equation}\label{alltherm}\end{subequations}
where $M_T$ is the pure thermal matrix in (\ref{purethermal}).

\subsection{Solutal problem}

\begin{figure}[t]
\centerline{
\begin{minipage}{8cm}
\psfig{file=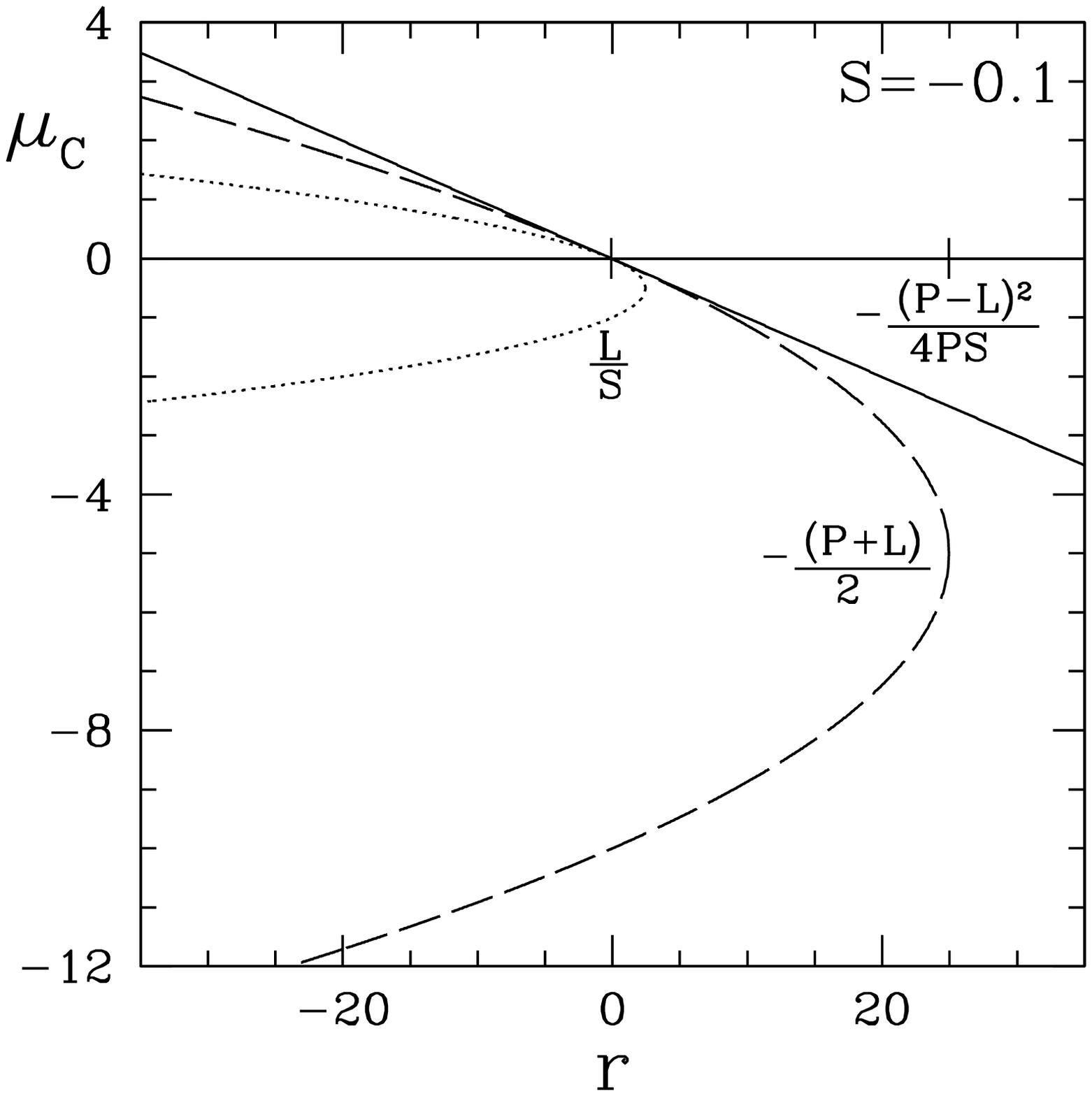,width=8cm}
\end{minipage}
\begin{minipage}{10cm}
\psfig{file=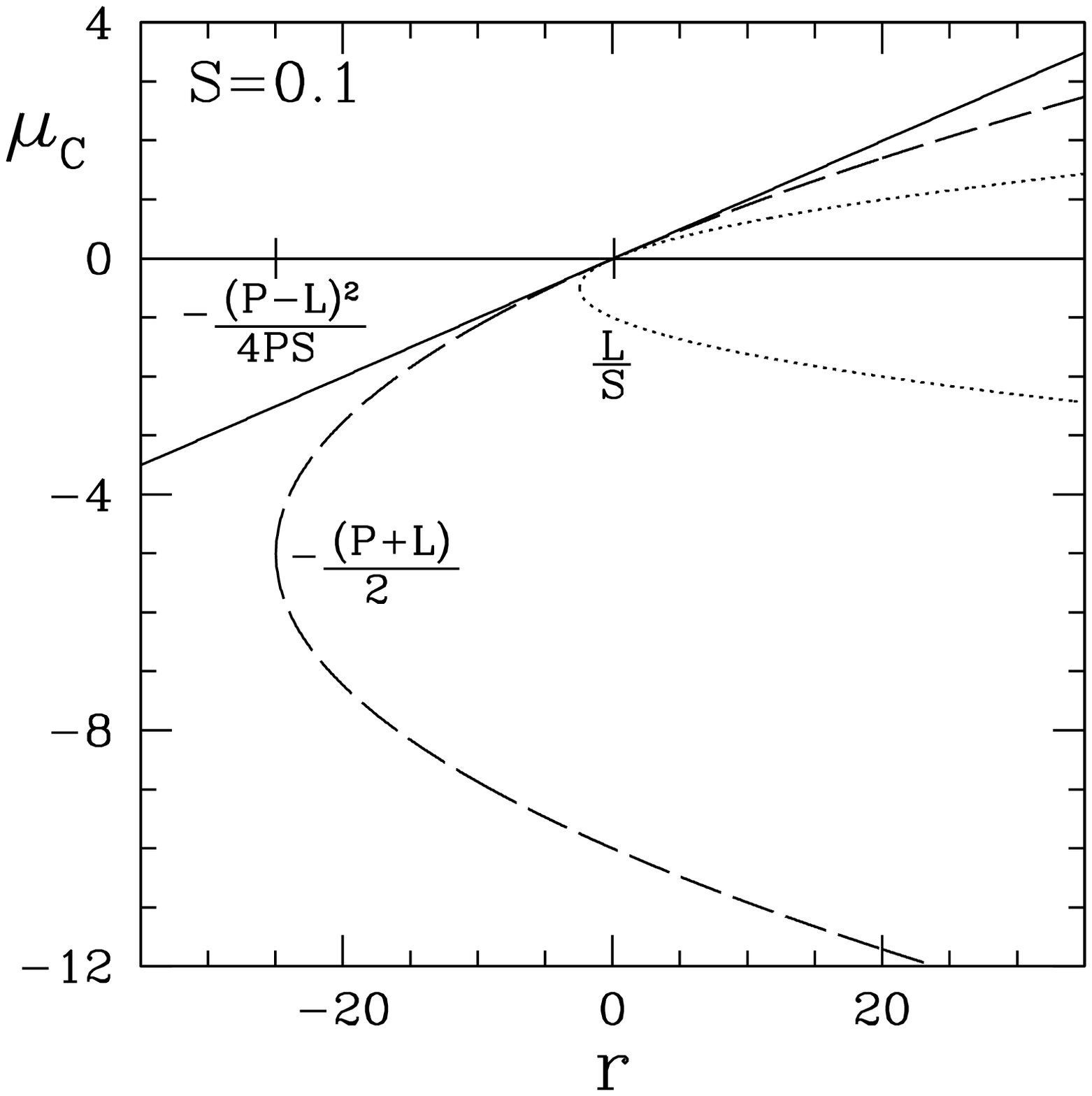,width=8cm}
\end{minipage}
}
\vspace*{-1cm}
\vspace*{0.5cm}
\caption{{} Eigenvalues of pure solutal problem for $L=0.01$.
and $S=-0.1$ (left) and $S=0.1$ (right).
Dotted and dashed curves shows $\mu_C$ for $P=1$ and $P=10$,
respectively. Solid curve shows $\mu_C$ for $P=100$, 
which is indistinguishable from $\sigmac$ valid for $P=\infty$.
Coordinates of vertex are shown for $P=10$.}
\label{fig:soleig}
\end{figure}

The pure solutal problem, described by the lower right 
$2 \times 2$ submatrix of (\ref{decmatrix}) is completely analogous 
to the thermal case,
with the inclusion of the Lewis number $L$ and the separation
parameter $S$:
\begin{equation}
\mu_C \left( \begin{array} {c} C \\ w_C \end{array} \right) 
=
\left( \begin{array} {c c}
-L & 1 \\
PSr & -P 
\end{array} \right)
\left( \begin{array} {c} C \\ w_C \end{array}
\right) 
\label{puresolutal}
\end{equation}

The solutal eigenvalues, plotted in figure \ref{fig:soleig}, are:
\begin{equation}
\mu_{C\pm} = \left[ -\left({{P+L}\over2} \right) \pm
\sqrt{\left({{P-L}\over2}\right)^2 + PSr} \right]
\label{puresolutaleigs}\end{equation}

We again require only the larger of the two eigenvalues 
$\mu_C \equiv \mu_{C+}$, which approaches $\sigmac$ for large $P$.
The onset of convection occurs at $\rc = L/S$, at which
$\mu_C$ has slope $PS/(P+L)$.
We again consider only the regime in which the 
eigenvalues are real, i.e. $((L-P)/2)^2 + PSr > 0$,

The corresponding normalized right and left solutal eigenvectors are:
\begin{subequations}\begin{eqnarray}
{\mathcal C}^R \equiv \left( \begin{array}{c} C^R \\ w_C^R \end{array} \right)
= {1\over {N_C}} \left( \begin{array}{c} 1 \\ L + \mu_C \end{array} \right)
\label{rightsol} \\
{\mathcal C}^L \equiv \left( \begin{array} {cc} C^L & w_C^L \end{array} \right)
= {{1}\over {N_C}} \left( \begin{array} {cc} P + \mu_C {\rm ~~}&{\rm ~~} 1 \end{array} \right)
\label{leftsol}
\end{eqnarray}
where $N_C^2 \equiv 2\sqrt{((P-L)/2)^2 + PSr}$
Thus
\begin{equation}
{\mathcal C}^L M_C {\mathcal C}^R = \mu_C
\label{diagsol}
\end{equation}\label{allsol}\end{subequations}
where $M_C$ is the pure solutal matrix in (\ref{puresolutal}).

\subsection{Thermosolutal coupling}
We now project the $4 \times 4$ thermosolutal problem
onto the most unstable thermal and solutal modes
to form the $2 \times 2$ matrix which constitutes our approximation.
We do so by multiplying the matrix of (\ref{decmatrix}) 
by left and right eigenvectors as follows:

\begin{subequations}\begin{eqnarray}
\left(\begin{array}{cc|cc} T^L & w_T^L & 0 & 0 \\ 
\hline 0 & 0 & C^L & w_C^L \end{array}\right)
&\left(\begin{array} {cc|cc}
-1 & 1 & 0 & 1 \\
Pr & -P & 0 & 0 \\
\hline
0 & 1 & -L & 1 \\
0 & 0 & PSr & -P
\end{array} \right)&
\left(\begin{array}{c|c} T^R & 0 \\ w_T^R & 0 \\ \hline 0 & C^R \\ 
0 & w_C^R \end{array}\right) 
\label{big1}\\
\nonumber\\
=\left(\begin{array}{cc} {\mathcal T}^L & 0 \\ 0 & {\mathcal C}^L \end{array}\right)
&\left(\begin{array} {cc} M_T & \Coupa \\ \Coupb & M_C \end{array}\right)&
\left(\begin{array}{cc} {\mathcal T}^R & 0 \\ 0 & {\mathcal C}^R \end{array}\right) 
\label{big2}\\
\nonumber\\
=&\left(\begin{array}{cc} 
{\mathcal T}^LM_T {\mathcal T}^R \; & \;{\mathcal T}^L \Coupa {\mathcal C}^R \\ 
{\mathcal C}^L \Coupb {\mathcal T}^R \; & \; {\mathcal C}^L M_C {\mathcal C}^R \end{array}\right)&
\label{big3}\\
\nonumber\\
=&\left(\begin{array}{cc} \mu_T & \coupa \\ 
\coupb & \mu_C \end{array}\right)&
\label{big4}
\end{eqnarray}\end{subequations}
In (\ref{big2})-(\ref{big3}), 
$M_T, M_C$ are the $2 \times 2$  pure thermal and solutal matrices and 
${\mathcal T}^L, {\mathcal T}^R, {\mathcal C}^L, {\mathcal C}^R$ the
corresponding left and right eigenvectors 
defined in (\ref{alltherm}) and (\ref{allsol}).
$\Coupa$, $\Coupb$ are the $2 \times 2$ off-diagonal submatrices 
in (\ref{big1}).
The calculation of the off-diagonal elements $\coupa$, $\coupb$ in (\ref{big4})
is tedious but straightforward. Their product is:
\begin{equation}
\coupa \coupb = {{P^2 S r^2}\over {4\sqrt{
((P-1)/2)^2 + Pr)((P-L)/2)^2 + PSr)}}}
\label{fTfC}
\end{equation}

\begin{figure}
\vspace*{-4cm}
\centerline{\psfig{file=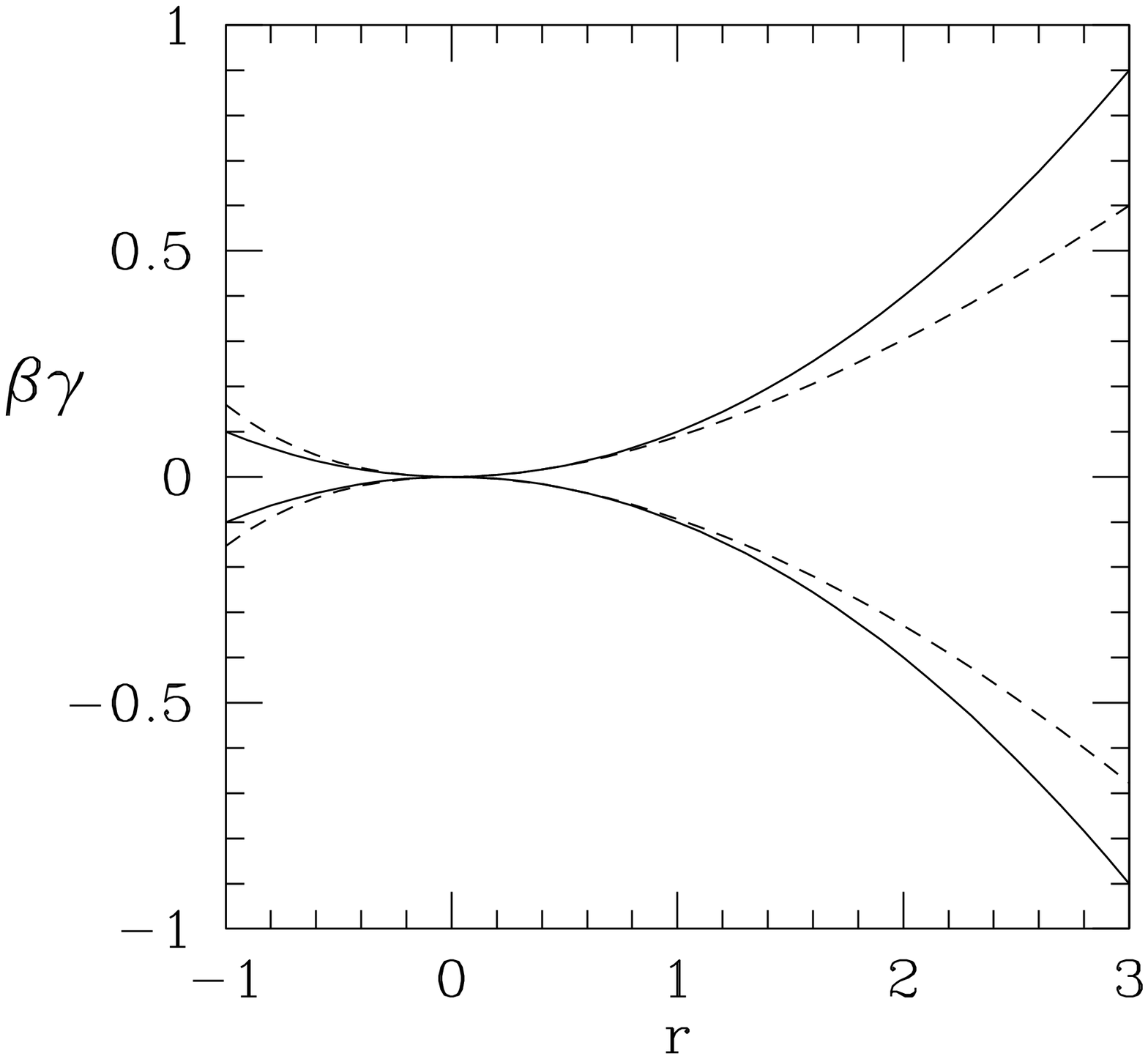,width=8cm}}
\vspace*{-1cm}
\caption{{} Product $\coupa\coupb$ of off-diagonal terms.
Dashed curves show expression (\ref{fTfC}) for $P=10$,
solid curves show $Sr^2$, valid for $P=\infty$.
Positive values correspond to $S=0.1$, negative to $S=-0.1$.}
\label{fig:offdiag}
\newpage
\centerline{
\begin{minipage}{9cm}
\psfig{file=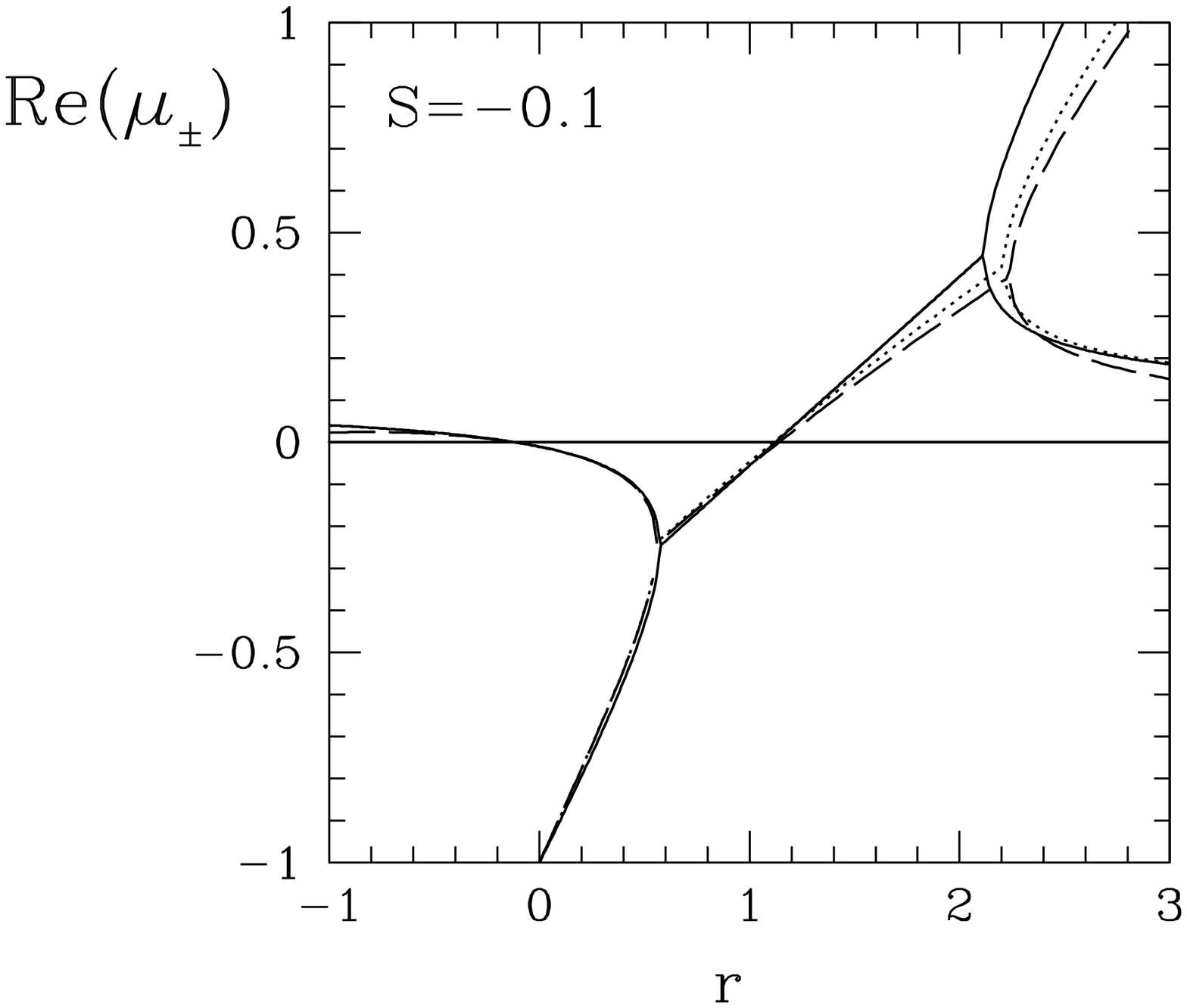,width=9cm}
\end{minipage}
\hspace*{-2cm}
\begin{minipage}{9cm}
\psfig{file=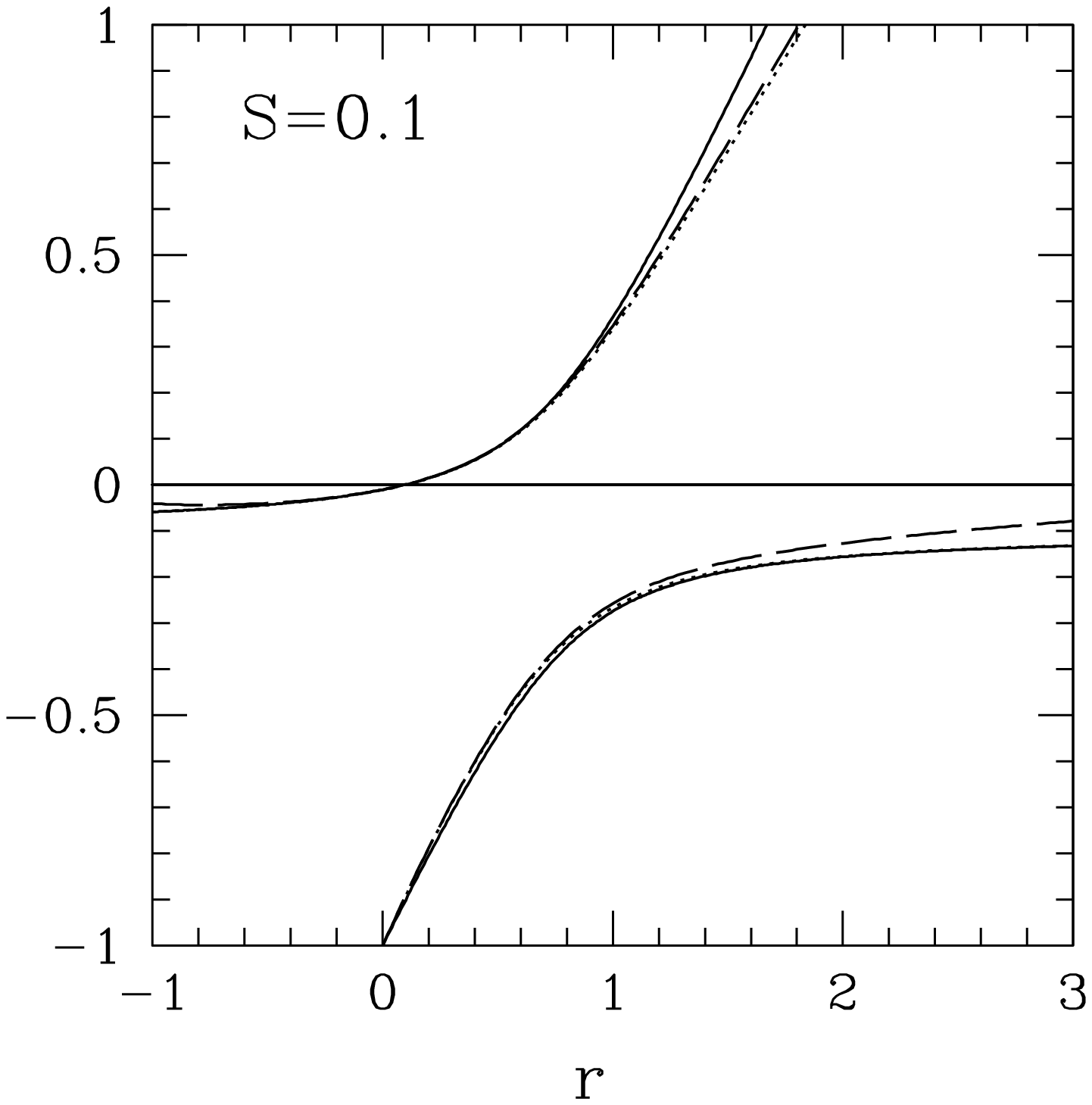,width=9cm}
\end{minipage}
}
\vspace*{-1.5cm}
\caption{{}
Real parts of the eigenvalues $\mu_\pm$ of the approximate 
$2\times 2$ matrix (dashed curves)
for $P=10$ and for $S=-0.1$ (left) and $S=0.1$ (right).
Shown for comparison are the real parts of the eigenvalues 
of the exact $3\times 3$ matrix (dotted curves) for $P=10$ and 
of the eigenvalues $\sigma_\pm$ (solid curves) for $P=\infty$.
}
\label{fig:2vs3eigs}
\end{figure}

This expression is not singular in the regime we consider here; 
the assumption that the eigenvalues $\mu_T$, $\mu_C$ of the thermal 
and the solutal problems be real requires that both factors
inside the square root be positive.
The coupling (\ref{fTfC}), plotted in figure \ref{fig:offdiag} for $P=10$, 
reduces to the far simpler coupling $Sr^2$ of (\ref{Sr2}) in the
limit of $P=\infty$ and shares its salient feature: its sign is that of $S$,
leading to avoided crossing if $S>0$ and complex coalescence if $S<0$.

\begin{figure}[h]
\vspace*{-1cm}
\centerline{
\psfig{file=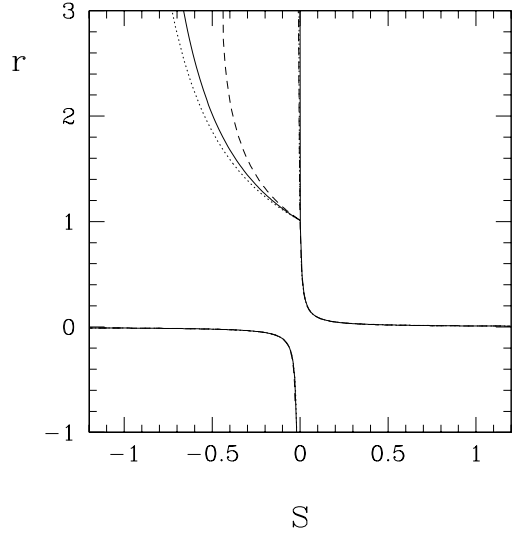,width=8cm}
}
\caption{{}
Thresholds for pitchfork and Hopf bifurcations.
Dashed curves are thresholds calculated from the approximate
$2\times 2$ matrix for $P=10$.
Dotted curves are thresholds calculated from the exact 
$3\times 3$ matrix for $P=10$.
Solid curves are thresholds for $P=\infty$.
The three thresholds for the pitchfork bifurcations cannot
be distinguished; all lie on solid curves.}
\label{fig:2vs3thresh}
\end{figure}

The eigenvalues of the $2\times 2$ approximate matrix are:
\begin{equation}
\mu_\pm = \frac{\mu_T+\mu_C}{2} \pm 
\sqrt{\left(\frac{\mu_T-\mu_C}{2}\right)^2 + \coupa\coupb}
\label{2x2eigs}\end{equation}
with $\mu_T=\mu_{T+}$, $\mu_C=\mu_{C+}$, and $\coupa\coupb$
given by (\ref{purethermaleigs}), (\ref{puresolutaleigs}), and (\ref{fTfC}).
In figures \ref{fig:2vs3eigs} and \ref{fig:2vs3thresh}, we compare
results from: \\
-- the $2\times 2$ approximate matrix (\ref{big4}) whose
eigenvalues $\mu_\pm$ are given by (\ref{2x2eigs}) for $P=10$ (dashed curves)\\
-- the $3\times 3$ exact matrix (\ref{thermosolutalmatrix}) for $P=10$
(dotted curves)\\
-- the $2\times 2$ matrix (\ref{lin2Dsys}) whose 
eigenvalues $\sigma_\pm$ are given by (\ref{2Deigs}) for $P=\infty$
(solid curves).

Specifically, in figure \ref{fig:2vs3eigs} we compare the real
parts of the eigenvalues (\ref{2x2eigs}) with those of (\ref{2Deigs}) 
and with those of the two eigenvalues of (\ref{thermosolutalmatrix})
with largest real part.
For the parameter values plotted, $P=10$ and $S=\pm 0.1$,
all three expressions give very similar results.
For both $S=\pm 0.1$, $\mu_+$ gives a slightly better approximation 
of the $3\times 3$ eigenvalue than $\sigma_+$, 
but $\sigma_-$ is slightly more accurate than $\mu_-$.
In figure \ref{fig:2vs3thresh}, we compare the thresholds for
pitchfork and Hopf bifurcations derived from these three matrices.
The thresholds for the pitchfork bifurcations are so close as
to be indistinguishable on the figure, whereas
the Hopf bifurcation threshold is overestimated by the 
$2\times 2$ approximate matrix.

Aside from these quantitative comparisons, figures
\ref{fig:2vs3eigs} and \ref{fig:2vs3thresh} demonstrate that
the exact finite Prandtl number linear stability problem
and our reduced model both exhibit the essential 
qualitative features of thermosolutal convection:
complex coalescence for $S<0$ and avoided crossing for $S>0$; 
pitchfork, Hopf, and codimension-two bifurcations.

\section{Conic sections and eigenvalues}
\label{Conic sections and eigenvalues}

Consider a $2\times 2$ matrix 
whose elements depend linearly on a parameter $r$:
\begin{eqnarray}
M &=& M_0 + r M_1 \nonumber \\
\left[\begin{array}{cc}
\alpha & \beta \\ \gamma & \delta \end{array}\right] &=&
\left[\begin{array}{cc} 
\alpha_0 & \beta_0 \\ \gamma_0 & \delta_0 \end{array} \right]
+ r \left[\begin{array}{cc} \alpha_1 & \beta_1 \\ \gamma_1 & \delta_1 
\end{array} \right]
\label{conic:matrix}\end{eqnarray}

We wish to describe the dependence of the eigenvalues of $M$ on $r$.
The equation obeyed by the eigenvalues $\lambda(r)$ is:
\begin{equation}
\lambda^2 - \Tr \lambda + \Det = 0
\label{conic:eigbasic}\end{equation}
Here, and throughout this appendix, $\Tr$ refers to the trace and not to
the temperature deviation; $\Det$ is the determinant of $M$. We have:
\begin{eqnarray}
\Tr &=& (\alpha_1  + \delta_1) \; r + \alpha_0 + \delta_0 
\nonumber \\
    &=& \Tr_1 r + \Tr_0     \label{conic:trace} 
\end{eqnarray}
\begin{subequations}\begin{eqnarray}
\Det     &=& (\alpha_1\delta_1 - \beta_1  \gamma_1)\; r^2 
       + (\alpha_0\delta_1+\alpha_1\delta_0-\beta_0\gamma_1 -\beta_1\gamma_0)\;r 
         + \alpha_0\delta_0-\beta_0  \gamma_0   \label{conic:dethalfdefine}\\
         &=& \Det_1 \; r^2 + 2 \; \Det_{1/2} \;r + \Det_0 \label{conic:det}
\end{eqnarray}\end{subequations}
where $\Tr_0$, $\Tr_1$, $\Det_0$, $\Det_1$
are the traces and determinants of the 
matrices $M_0$ and $M_1$ in (\ref{conic:matrix}) and 
$\Det_{1/2}$ is defined by (\ref{conic:dethalfdefine})-(\ref{conic:det}).
The eigenvalues of $M$ are complex where the discriminant $\Disc$ is negative, where
\begin{subequations}\begin{eqnarray}
\Disc \equiv \Tr^2-4\Det  
&=& (\Tr_1^2-4 \Det_1)\;r^2+2(\Tr_0 \Tr_1 - 4\Det_{1/2})\;r + (\Tr_0^2-4 \Det_0) 
\label{discc}\\
&=& \Disc_1 \; r^2 + 2 \Disc_{1/2} \; r + \Disc_0 \label{discd}
\end{eqnarray}\end{subequations}
Here $\Disc_0$, $\Disc_1$ are the discriminants of $M_0$, $M_1$ and 
$\Disc_{1/2}$ is defined by (\ref{discc})-(\ref{discd}).
Whether and where $\Disc$ is negative is in turn seen from (\ref{discd})
to be determined by $\Disc_1$ and by 
\begin{equation}
\Sgen \equiv \frac{\Disc_{1/2}^2 - \Disc_1 \Disc_0}{4 \Disc_1}.
\label{conic:Sgendefine}\end{equation}
$\Disc_1$ and $\Sgen$ are both {\it invariants} under translation and rotation of
$(r,\lambda)$ of equation (\ref{conic:eigbasic}); see, e.g., \cite{Ayre}.

Writing $\lambda=\sigma + i \omega$, substituting (\ref{conic:trace}) and
(\ref{conic:det}) into (\ref{conic:eigbasic}), and separating
into real and imaginary parts, we obtain:
\begin{subequations}\begin{eqnarray}
\sigma^2 -\omega^2 -\Tr_1\sigma r +\Det_1r^2 -\Tr_0\sigma +2\Det_{1/2}r +\Det_0 &=& 0
\label{conic:real}\\
(2\sigma - \Tr_1 r - \Tr_0) \; \omega &=& 0 \label{conic:imag}
\end{eqnarray}\end{subequations}
According to equation (\ref{conic:imag}), 
\begin{subequations}\begin{eqnarray}
\mbox{either   } && \omega=0 \label{conic:noimag}\\
\mbox{or    } && \sigma=\frac{1}{2}(\Tr_0 + \Tr_1 r) \label{conic:realavg}
\end{eqnarray}\end{subequations}
If $\omega=0$, then (\ref{conic:real}) becomes: 
\begin{equation}
\sigma^2 - \Tr_1 \sigma r + \Det_1 r^2 - \Tr_0\sigma + 2\Det_{1/2}r +\Det_0 = 0
\label{conic:sigma}
\end{equation}
If $\omega\neq 0$, then substituting (\ref{conic:realavg}) 
into (\ref{conic:real}) yields:
\begin{equation}
\omega^2 +\frac{\Disc_1}{4} r^2 + \frac{\Disc_{1/2}}{2} r 
+ \frac{\Disc_0}{4} = 0
\label{conic:omega}\end{equation}

The qualitative nature of the solutions to 
the second-degree equations (\ref{conic:sigma}) 
and(\ref{conic:omega}) depends on the sign of 
$\Disc_1$ and $\Delta$.
If the discriminant $\Disc_1$ is positive, 
then the set $(r,\sigma)$ satisfying (\ref{conic:sigma})
is a hyperbola and the set $(r,\omega)$ satisfying (\ref{conic:omega}) 
is an ellipse, and vice versa if $\Disc_1$ is negative.
If $\Disc_1$ vanishes, then both sets are parabolas.
In almost all the cases we shall study, 
$\Disc_1$ is positive and we shall assume this from now on.

We rewrite equation (\ref{conic:omega}) as:
\begin{subequations}\begin{eqnarray}
\omega^2 &+& \frac{\Disc_1}{4} \left(r + \frac{\Disc_{1/2}}{\Disc_1} \right)^2
       = -\frac{1}{4\Disc_1}\left(\Disc_0\Disc_1 - \Disc_{1/2}^2 \right) 
\label{conic:radefine}\\
\omega^2 &+& \frac{\Disc_1}{4} (r-\rmid)^2 = -\Sgen
\label{conic:radefine2}\end{eqnarray}\label{conic:ellipse}\end{subequations}
where $\Sgen$ is defined by (\ref{conic:Sgendefine}) and $\rmid$ by
(\ref{conic:radefine})-(\ref{conic:radefine2}).
The sign of $\Sgen$ is also crucial: if $\Sgen>0$, then 
(\ref{conic:ellipse}) describes a degenerate ellipse containing no points:
there are no complex eigenvalues. If $\Sgen=0$, then the
ellipse contains just the point $(\rmid,0)$.
If $\Sgen<0$, then (\ref{conic:ellipse}) describes an ellipse
whose two semiaxes are $r=\rmid$ of length $\sqrt{-\Sgen}$
and $\omega=0$ of length $\sqrt{-4\Sgen/\Disc_1}$;
complex eigenvalues exist over the interval
\begin{equation}
 \vert r-\rmid \vert < \sqrt{-4\Sgen/\Disc_1}
\label{conic:complexrange} \end{equation}
At $r=\rmid$, $\omega$ attains its maximum value of $\sqrt{-\Sgen}$ and
(\ref{conic:realavg}) implies $\sigma(\rmid)=\sigmamid$, where
\begin{equation}
\sigmamid \equiv 
\frac{1}{2}\left(\Tr_0  - \Tr_1 \frac{\Disc_{1/2}}{\Disc_1} \right) =
\frac{\Tr_0 \Disc_1 - \Tr_1\Disc_{1/2}}{2\Disc_1}
=\frac{-2 \Tr_0 \Det_1 + 2 \Tr_1 \Det_{1/2}}{\Disc_1}
\label{conic:sigmidefine}
\end{equation}

We now turn to the hyperbola described by (\ref{conic:sigma})
when $\Disc_1 > 0$; in particular we seek to characterize it by its asymptotes.
The first three terms of (\ref{conic:sigma}) 
imply that the sum of the slopes of the asymptotes 
is $\Tr_1$ and their product is $\Det_1$, i.e. the slopes are
the eigenvalues $\lambda_{1\pm} = (\Tr_1 \pm \sqrt{\Disc_1})/2$ of $M_1$.
These are real and distinct by the assumption $\Disc_1 > 0$.
Some more algebra shows that (\ref{conic:sigma}) is equivalent to:
\begin{eqnarray}
\left(\sigma - \sigmamid - \lambda_{1+}(r -\rmid)\right)  
\left(\sigma - \sigmamid - \lambda_{1-}(r -\rmid)\right) = \Sgen \label{conic:hyperbola}
\end{eqnarray}
where $\rmid$, $\Sgen$, and $\sigmamid$ are defined in 
(\ref{conic:radefine})-(\ref{conic:radefine2}) and (\ref{conic:sigmidefine})

The asymptotes are the roots of the two factors in (\ref{conic:hyperbola}).
They intersect at $(\rmid,\sigmamid)$ and divide the plane into four quadrants.
The magnitude of $\Sgen$ measures the distance of closest approach of
the two portions of the hyperbola; its sign 
determines which two of the four quadrants are occupied by the hyperbola. 
If $\Sgen > 0$, then one branch $\sigma_+$ of the hyperbola lies above 
both asymptotes (in the sense of greater $\sigma$) 
and the other branch $\sigma_-$ lies below them.
Each branch exists for all $r$.
This is the situation called {\it avoided crossing}.
When $\Sgen = 0$, the hyperbola is degenerate and consists precisely of the 
two intersecting asymptotic lines. In this case, the two branches 
$\sigma_+$ and $\sigma_-$ can be considered to either behave
non-smoothly or to exchange identities at $r=\rmid$.
If $\Sgen<0$, then both values $\sigma_{\pm}$ lie between the asymptotes,
i.e. they both lie above one asymptote and below the other.
In this case, there are no real solutions to (\ref{conic:hyperbola}) 
in the range (\ref{conic:complexrange}) surrounding $\rmid$.
At the endpoints of the interval in (\ref{conic:complexrange}),
the curves $\sigma_{\pm}$ join, to be replaced within this interval 
by the single linear segment (\ref{conic:realavg}).
This is the situation we call {\it complex coalescence}.

The phenomenon of avoided crossing can be quantified by
differentiating (\ref{conic:hyperbola}) implicitly with respect to $r$:
\begin{subequations}\begin{eqnarray}
(\sigma' - \lambda_{1+})(\sigma - \sigmamid - \lambda_{1-}(r -\rmid))
&+&(\sigma' - \lambda_{1-})(\sigma - \sigmamid - \lambda_{1+}(r -\rmid))=0 
\;\;\;\;\;\; \\
\sigma''(2(\sigma - \sigmamid) - (\lambda_{1-}+\lambda_{1+})(r -\rmid))
&+&2(\sigma' - \lambda_{1-})(\sigma' - \lambda_{1+}) = 0\\
\sigma'''(2(\sigma - \sigmamid) - (\lambda_{1-}+\lambda_{1+})(r -\rmid))
&+&3\sigma''(2\sigma' - (\lambda_{1-}+\lambda_{1+})) = 0
\end{eqnarray}\end{subequations}
and then evaluating $\sigma$, $\sigma'$, $\sigma''$, and $\sigma'''$ 
successively at $\rmid$:
\begin{subequations}\begin{eqnarray}
\sigma_\pm &=& \sigmamid \pm \sqrt{\Sgen} \\
\sigma_\pm' &=& \frac{\lambda_{1+}+\lambda_{1-}}{2} \\
\sigma_\pm'' &=& \frac{(\lambda_{1+}-\lambda_{1-})^2}{\pm 4\sqrt{\Sgen}} \\
\sigma_\pm''' &=& 0
\end{eqnarray}\label{derivs}\end{subequations}
We see from (\ref{derivs}) that the change in slope undergone by $\sigma_\pm$
at $\rmid$ corresponds to an extremum in $\sigma_\pm''$,
whose magnitude measures the abruptness of the change.

Another fact which we shall use is that
a line in the $(r,\sigma)$ plane which is parallel, but not equal, 
to one of the asymptotes intersects the hyperbola in exactly one point,

We briefly discuss the exceptional case $\Disc_1=0$.
Equation (\ref{conic:omega}) for the imaginary part of the eigenvalues
becomes:
\begin{equation}
0 = \omega^2 + \frac{\Disc_{1/2}}{2} r + \frac{\Disc_0}{4}
\label{conic:omega_parabola}
\end{equation}
Equation (\ref{conic:sigma}) for the real part of the eigenvalues becomes:
\begin{eqnarray}
0 &=& \left(\sigma - \frac{\Tr_1}{2} r -\frac{\Tr_0}{2}\right)^2 
+ \left(2\Det_{1/2}-\frac{\Tr_0 \Tr_1}{2} \right)r 
+ \Det_0 - \frac{\Tr_0^2}{4} \nonumber\\
&=&\left(\sigma - \frac{\Tr_1}{2} r -\frac{\Tr_0}{2}\right)^2 
- \frac{\Disc_{1/2}}{2} r - \frac{\Disc_0}{4}
\label{conic:sigma_parabola}
\end{eqnarray}
Both (\ref{conic:omega_parabola}) and (\ref{conic:sigma_parabola}) 
describe parabolas.
The parabola of (\ref{conic:omega_parabola}) is oriented along the
the $r$-axis, while the axis of (\ref{conic:sigma_parabola}) 
is the line $\sigma = (\Tr_1 r + \Tr_0)/2$.
If $\Tr_1=0$, then (\ref{conic:sigma_parabola}) is
also oriented along the $r$-axis.
The two parabolas are oriented in opposite directions:
(\ref{conic:omega_parabola}) opens towards positive values of $r$
if $\Disc_{1/2}$ is negative and vice versa for (\ref{conic:sigma_parabola}).
The vertex of both parabolas is located at $r=-\Disc_0/(2\Disc_{1/2})$,
with $\omega = 0$ for (\ref{conic:omega_parabola}) and
$\sigma = (-\Disc_0\Tr_1/(2\Disc_{1/2}) + \Tr_0)/2$
for (\ref{conic:sigma_parabola}).

Our treatment of binary fluid convection leads to
two matrices of type (\ref{conic:matrix}),
one whose eigenvalues $\sigma$ govern the linear stability of the conductive state
and the other whose eigenvalues $E$ are the kinetic energy of nonlinear steady states.
$L$ is the Lewis number, which is necessarily positive and usually small,
$S$ is the separation parameter, which may have either sign, and
$r$ is the reduced Rayleigh number.
In tables 1 and 2, we give the quantities we have defined above for each 
of these two matrices.

\begin{eqnarray}
\begin{array} {|l|l|l|} \hline
\multicolumn{3}{|c|} {\mbox{Growth rate matrix}}    \\ \hline
\multicolumn{3}{|c|} 
{\left(\begin{array}{cc} r-1 & r S \\ r & r S - L\end{array}\right) =
\left(\begin{array}{cc} -1 & 0 \\ 0 & -L  \end{array}\right) 
+r\left(\begin{array}{cc} 1 & S \\ 1 & S \end{array}\right)} \\ \hline
\Tr_0=-(1+L)   & \Det_0=L          & \Disc_0=(1-L)^2 \\
               & 2\;\Det_{1/2}=-(S+L) & \Disc_{1/2}=-(1-L)(1-S) \\
\Tr_1=1+S      & \Det_1=0          & \Disc_1 = (1+S)^2 \\ \hline
\multicolumn{3}{|c|} {\Sgen = S\dfrac{(1-L)^2}{(1+S)^2}}    \\ \hline
\multicolumn{3}{|c|} {\rmid = \dfrac{(1-L)(1-S)}{(1+S)^2}}  \\ \hline
\multicolumn{3}{|c|} {\sigmamid = -\dfrac{S+L}{1+S}}        \\ \hline
\multicolumn{3}{|c|} {\mbox{ellipse: }
\omega^2 + \dfrac{(1+S)^2}{4} \left(r - \dfrac{(1-L)(1-S)}{(1+S)^2} \right)^2
       = -S \dfrac{(1-L)^2}{(1+S)^2}} \\ \hline
\multicolumn{3}{|c|} {\mbox{hyperbola: }
\left(\sigma + \dfrac{S+L}{1+S} - (1+S)\left(r - \dfrac{(1-L)(1-S)}{(1+S)^2}\right)\right) 
\left(\sigma + \dfrac{S+L}{1+S}\right) = S\;\dfrac{(1-L)^2}{(1+S)^2}} \\ \hline
\end{array}
\nonumber\end{eqnarray}
\begin{eqnarray}
\begin{array} {|l|l|l|} \hline
\multicolumn{3}{|c|} {\mbox{Energy matrix}}    \\ \hline
\multicolumn{3}{|c|} 
{\left(\begin{array}{cc} r-1 & Sr \\ Lr & L(Sr-L) \end{array}\right) 
= \left(\begin{array}{cc} -1 & 0 \\ 0 & -L^2 \end{array}\right) 
+r \left(\begin{array}{cc} 1 & S \\ L & LS \end{array}\right)} \\ \hline
\Tr_0=-(1+L^2) & \Det_0=L^2        & \Disc_0=(1-L^2)^2 \\
               & 2\;\Det_{1/2}=-L(S+L)& \Disc_{1/2}=-(1-L^2)(1-LS) \\
\Tr_1=1+LS     & \Det_1=0          & \Disc_1 = (1+LS)^2     \\ \hline
\multicolumn{3}{|c|} {\tilde{\Sgen} = SL\dfrac{(1-L^2)^2}{(1+LS)^2}}   \\ \hline
\multicolumn{3}{|c|} {\rmidnl = \dfrac{(1-L^2)(1-LS)}{(1+LS)^2}} \\ \hline
\multicolumn{3}{|c|} {\Emid = -\dfrac{L(S+L)}{1+LS}}       \\ \hline
\multicolumn{3}{|c|} {\mbox{ellipse: }
\omega^2 + \dfrac{(1+LS)^2}{4} \left(r - \dfrac{(1-L^2)(1-LS)}{(1+LS)^2} \right)^2
       = -LS \dfrac{(1-L^2)^2}{(1+LS)^2}} \\ \hline
\multicolumn{3}{|c|} {\mbox{hyperbola: }
\left(E + \dfrac{L(S+L)}{1+LS} - (1+LS)\left(r - \dfrac{(1-L^2)(1-LS)}{(1+LS)^2}\right)\right) 
\left(E + \dfrac{L(S+L)}{1+LS}\right) = SL\;\dfrac{(1-L^2)^2}{(1+LS)^2}} \\ \hline
\end{array}
\nonumber\end{eqnarray}
\newpage
\bibliography{binary}
\end{document}